\documentstyle[preprint,epsfig,aps,floats,prc]{revtex}  

\begin{document}                
%
%

\title{$\phi$ and $\omega$ Meson Production in pp Reactions
        at $p_{lab}$=3.67 GeV/c}
\author{
  F.~Balestra$^{1}$, Y.~Bedfer$^{2,a}$, R.~Bertini$^{1,2}$, L.C.~Bland$^3$,
  A.~Brenschede$^{4,c}$, F.~Brochard$^{2,b}$, M.P.~Bussa$^1$,  
  Seonho~Choi$^{3,d}$, M.~Debowski$^5$, R.~Dressler$^6$, M.~Dzemidzic$^{3,e}$,
  J.-Cl.~Faivre$^{2,a}$, I.V.~Falomkin$^{7,f}$, L.~Fava$^8$, L.~Ferrero$^1$,
  J.~Foryciarz$^{9,5,g}$, I.~Fr\"ohlich$^4$, V.~Frolov$^7$, R.~Garfagnini$^1$,
  A.~Grasso$^1$, E.~Grosse$^6$, S.~Heinz$^{1,2}$, V.V.~Ivanov$^7$,
  W.W.~Jacobs$^3$, W.~K\"uhn$^4$, A.~Maggiora$^1$, M.~Maggiora$^1$, 
  A.~Manara$^{1,2}$, D.~Panzieri$^8$, H.-W.~Pfaff$^4$, G.~Piragino$^1$,
  G.B.~Pontecorvo$^7$, A.~Popov$^7$, J.~Ritman$^4$, P.~Salabura$^5$,
  V.~Tchalyshev$^7$, F.~Tosello$^1$, S.E.~Vigdor$^3$, and G.~Zosi$^1$}

\address{
(DISTO Collaboration) \\
$^1$Dipartimento di Fisica ``A. Avogadro'' and INFN - Torino, Italy  \\
$^2$Laboratoire National Saturne, CEA Saclay, France\\
$^3$Indiana University Cyclotron Facility, Bloomington, Indiana, U.S.A.\\
$^4$II. Physikalisches Institut,  University of Gie\ss en, Germany \\
$^5$M. Smoluchowski Institute of Physics, Jagellonian University, Krak\'ow, 
Poland \\
$^6$Forschungszentrum Rossendorf, Germany \\
$^7$JINR, Dubna, Russia \\
$^8$Universita' del Piemonte Orientale and INFN - Torino, Italy \\
$^9$H.Niewodniczanski Institute of Nuclear Physics, Krak\'ow, Poland \\
}

\date{\today}     
\maketitle 
\begin{abstract}                
  The exclusive production cross sections for $\omega$ and $\phi$ mesons have
  been measured in proton-proton reactions at $p_{lab}=3.67$~GeV/c.  The
  observed $\phi/\omega$ cross section ratio is
  $(3.8\pm0.2^{+1.2}_{-0.9})\times 10^{-3}$.  After phase space corrections,
  this ratio is enhanced by about an order of magnitude relative to naive
  predictions based upon the Okubo-Zweig-Iizuka (OZI) rule, in comparison to
  an enhancement by a factor $\sim 3$ previously observed at higher beam
  momenta. The modest increase of this enhancement near the production
  threshold is compared to the much larger increase of the $\phi/\omega$ ratio
  observed in specific channels of $\bar pp$ annihilation experiments.
  Furthermore, differential cross section results are also presented which
  indicate that although the $\phi$ meson is predominantly produced from a
  $^3P_1$ proton-proton entrance channel, other partial waves
  contribute significantly to the production mechanism at this beam
  momentum.

\end{abstract}
\pacs{PACS numbers: 25.40.Ve, 13.75.Cs, 13.85.Hd, 14.40.Cs}

\section{Introduction}                          \label{sec:intro}

The proton is a complex composite object in which sea-quarks and gluons make
significant contribution to the structure functions.  Investigation of the
strange structure of the nucleon is currently of great interest since strange
quarks appear only as sea quarks in most models of the nucleon in contrast to
the light quarks ($u$ and $d$) that appear as both valence and sea quarks.

An important experimental approach to study the role of strange quarks in the
proton's wave function is to measure the relative production of $\phi$ and
$\omega$ vector mesons.  Since the singlet-octet mixing angle of the ground
state vector meson nonet is close to the ideal value (i.e.
$\tan\theta_{ideal}=1/\sqrt{2}$)~\cite{Cas98}, the $\omega$ meson consists
almost completely of light valence quarks and the $\phi$ meson almost
completely of strange valence quarks.  According to the Okubo-Zweig-Iizuka
(OZI) rule, processes with disconnected quark lines in the initial or final
state are suppressed~\cite{Zwe64,Oku65,Iiz66,Oku77}.  As a result, the
production of the $\phi$ meson is expected to be strongly suppressed relative
to the $\omega$ meson in hadronic reactions with no strange quarks in the
initial state.  Under the assumption that the OZI rule is exactly fulfilled,
the $\phi$ meson can only be produced by the small admixture of light quarks
to its wave function.  Thus, the relative production cross sections for the
$\phi$ and $\omega$ mesons ($R_{\phi/\omega}$) can be calculated by the
following formula:
\begin{equation}
R_{\phi/\omega} = f \times \tan^2 \delta = f \times 4.2\cdot10^{-3}
\label{Eq:OZI}
\end{equation}
where $f$ is a correction for the available phase space, and
$\delta\approx3.7^\circ$ is the deviation from the ideal mixing 
angle~\cite{Lip76}.

In certain hadronic reactions it has been observed that the ratio of the
exclusive cross sections for the $\phi$ and $\omega$ meson production
reactions significantly exceeds estimates based on simple quark models (see
e.g.~\cite{Ell95}).  For $\bar pp$ annihilation at rest this enhancement is
observed to be particularly dramatic~\cite{Rei91,Abl94,Ams95}.  This apparent
violation of the OZI rule has been interpreted as evidence for a
non-negligible negatively polarized $s \bar s$ Fock component to the proton's
wave function~\cite{Ell95,Ell99}.  Here it is important to distinguish between
extrinsic and intrinsic quark and gluon contributions to the nucleon
sea~\cite{Bro96,Bur92}, since other explanations stress the importance of
higher order rescattering
processes~\cite{Lip84,Mei97,Mar99,Nak99,Sib96,Tit99}, thereby "avoiding" the
OZI rule.

Recently, there have been numerous experimental investigations on the
contribution of strange quarks to the proton: At low momentum transfers $q^2$
there are measurements of parity violating polarized electron-proton
scattering to determine the strange electromagnetic form factors of the
nucleon.  First results indicate very small effects~\cite{Ani99,Spa00}, and
thus the scale of the strange vector current $\langle N|\bar s \gamma_\mu
s|N\rangle$ can not yet be determined (see e.g.~\cite{Ham99}).  Spin dependent
structure functions have also been measured in a number of
inclusive~\cite{Ash88,Ash89,Ama92,ant93,abe95,Ad95,Abe97,ad97,Air98} and
semi-inclusive~\cite{Ade98,Ack99} experiments on deep inelastic scattering of
polarized leptons from nucleons, suggesting that the strange sea quarks are
weakly polarized against the proton helicity.  Furthermore, it has commonly
been discussed that the magnitude of the $\Sigma_{\pi N}$ term in low energy
pion nucleon elastic scattering~\cite{pns,Che76,Gas88,Gas91} indicates a
possibly sizable, but very uncertain scalar density of strange sea quarks
$\langle N|\bar ss|N\rangle$ in the nucleon's wave function.

The negatively polarized intrinsic strangeness model was in part motivated by
the strong correlation of the $\phi$ meson yield to the spin triplet fraction
of the annihilating $p\bar p$ system~\cite{Ber96}.  Since $\phi$ meson
production in $pp$ reactions at threshold must proceed via the spin triplet
entrance channel, further insight on the origin of the enhanced $\phi$
production in $p\bar p$ could be provided by studying near-threshold $pp$
reactions, where predictions based upon the strange sea quarks in the nucleon
and meson rescattering models might be expected to differ.

To address this problem we have measured the production of $\phi$ and $\omega$
mesons in proton-proton reactions with the DISTO spectrometer~\cite{disto} at
the Saturne accelerator in Saclay.  We present here total and differential
cross sections that were determined for a beam momentum of $3.67$~GeV/c which
corresponds to a much lower available energy above the $\phi$ production
threshold ($Q=\sqrt{s}-\sqrt{s_0}=83$~MeV) than the existing data
($Q>1.6$~GeV)~\cite{Blo75,Bal77,Are82,Gol97}.  In previous
letters~\cite{Bal98,Bal99} we have presented the first results close to the
$\phi$ meson production threshold in proton-proton reactions, where it was
determined that the $\phi/\omega$ cross section ratio increases slightly
toward threshold.  Since these publications we have made several improvements
to the data analysis, including refinements of the acceptance correction,
improvements in the tracking algorithm and calibrations, as well as the
analysis of a factor four more statistics. These results are presented here in
more detail, together with differential cross section distributions that were
previously not available.  All of the previously published data are consistent
with the results presented here.

This paper is organized as follows: in the next section experimental
details of the detector and data analysis are presented.  The resulting $\phi$
and $\omega$ meson total and differential cross sections are presented in the
ensuing section, followed by a discussion of these results and a summary.

\section{Experiment}              \label{sec:exp}
  \subsection{Apparatus}                        \label{ssec:appar}   
\begin{figure}
  \begin{center}
   \mbox{\epsfig{figure={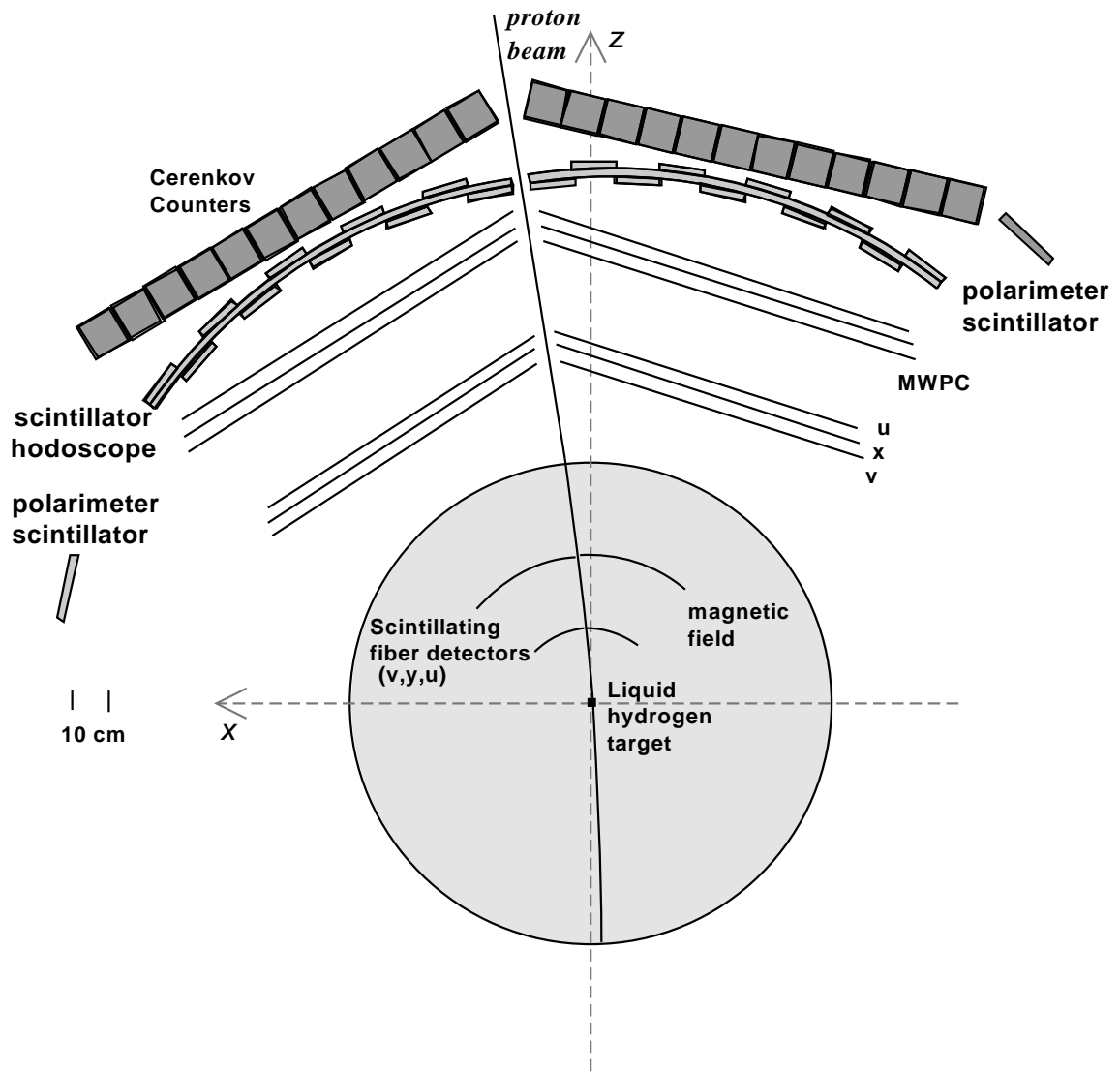},
        width=0.97\linewidth }}
  \end{center}
  \caption[The DISTO Spectrometer.]
  {\label{Fig:DISTO} Schematic layout of the DISTO experimental apparatus,
    viewed from above. The large shaded area represents the effective field
    region.  }
\end{figure}

  A proton beam from the SATURNE synchrotron with momentum
  $p_{lab}=3.67$~GeV/c was directed onto a liquid hydrogen target of 2~cm
  length, and multiple charge particle final states were measured with the
  DISTO spectrometer~\cite{disto}, which is shown schematically in
  Figure~\ref{Fig:DISTO}.  Charged particles were tracked through a magnetic
  spectrometer and detected by a scintillator hodoscope and an array of water
  \v Cerenkov detectors. The magnetic spectro\-meter consisted of a dipole
  magnet (1.0~T$\cdot$m), 2 sets of scintillating fiber hodoscopes inside the
  field, and 2 sets of multi wire proportional chambers (MWPC) outside the
  field. The large acceptance of the spectrometer ($\approx\pm 15^\circ$
  vertical, $\pm 48^\circ$ horizontal) allowed for coincident detection of
  four charged particles, which was essential for the kinematically complete
  reconstruction of many final states ($pp\pi^+\pi^-$, $pp\pi^+\pi^-\pi^0$,
  $ppK^+K^-$, $pK\Lambda$, $pK\Sigma$).  Particle identification and
  4-momentum conservation served as powerful tools for background rejection.
  Event readout was triggered by a multiplicity condition on the scintillating
  fiber and hodoscope detectors, selecting events with at least three charged
  particles.  Details of the experimental apparatus can be found in
  Ref.~\cite{disto}.

  \subsection{Data Analysis}                    \label{ssec:dana}
  The exclusive reactions $pp\rightarrow pp\eta$ and $pp\rightarrow pp\omega$
  were identified via the $\pi^+\pi^-\pi^0$ decay of the $\eta$ and $\omega$
  mesons with partial widths $\Gamma/\Gamma_{tot}= 0.232$ and
  $\Gamma/\Gamma_{tot}= 0.888$, respectively~\cite{Cas98}. For the
  $pp\rightarrow pp\phi$ reaction, the $\phi$ was observed via its $K^+K^-$
  decay with $\Gamma/\Gamma_{tot}=0.491$.  Since these channels all have four
  charged particles in the final state (three positive and one negative),
  these events could be reconstructed by applying particle identification and
  kinematical constraints on an event-by-event basis to the same sample of
  four track events.

    \subsubsection{Tracking}
    After calibration, the data from the magnetic spectrometer consist of a
    collection of points in space determined by the various detector
    components.  Thus, the first major step in the data analysis is to
    determine the trajectories of the charged particles from the detector
    position information.  In general, a particle's trajectory is defined by 5
    independent parameters.  These parameters can be chosen in numerous ways,
    however they must span the space of allowed possibilities.  The choice of
    track parameters used in this experiment to define the trajectories was
    motivated by the dipole magnetic field geometry, and was based on a right
    handed coordinate system with its origin in the center of the target,
    $\hat z$ in the beam direction at the target, and $\hat y$ (vertical)
    along the magnetic field direction.  The five track parameters used were
    $X$ and $Y$ to describe the intersection of the track with the $\hat
    x-\hat y$ plane through the origin, $\phi$ to describe the angle of the
    track's projection onto the $\hat x-\hat z$ plane at the point $(X,Y,0)$,
    $m$ to describe the ratio of the particle's momentum along $\hat y$ to the
    momentum in the $\hat x-\hat z$ plane, and $p_{xz}$ for the reciprocal of
    the momentum in the $\hat x-\hat z$ plane.

    The observed positions in the
    detectors can be calculated as a function of these
    five parameters by detailed Monte-Carlo simulations of the detector
    performance.  The goal of the tracking algorithm is to invert this
    function, i.e. determine the track parameters from the measured position
    information.  This is done in a two step procedure: In the first step
    (track-search) it is determined which hits belong together to form a
    track. Then, in the track-fit procedure the five track parameters are
    determined from the hits in an iterative, quasi-Newton procedure by
    interpolating within a matrix of reference trajectories.

    \subsubsection{Particle Identification}    
    Particle identification was achieved using the light output from the water
    \v Cerenkov detectors, which provided good $\pi^+$ - proton separation
    spanning a wide range of momenta, as well as $K^\pm$ identification in a
    restricted range above the kaon \v Cerenkov threshold $p_{K,th}~=~M_Kc /
    \sqrt{n^2-1}$.  Water was chosen as the \v Cerenkov radiator ($n=1.33$,
    $p_{K,th}=560$~MeV/c) in order to match the momentum range of the kaons
    from the reaction $pp\rightarrow ppK^+K^-$, which are distributed around
    700~MeV/c.
    
\begin{figure}
  \begin{center}
    \epsfig{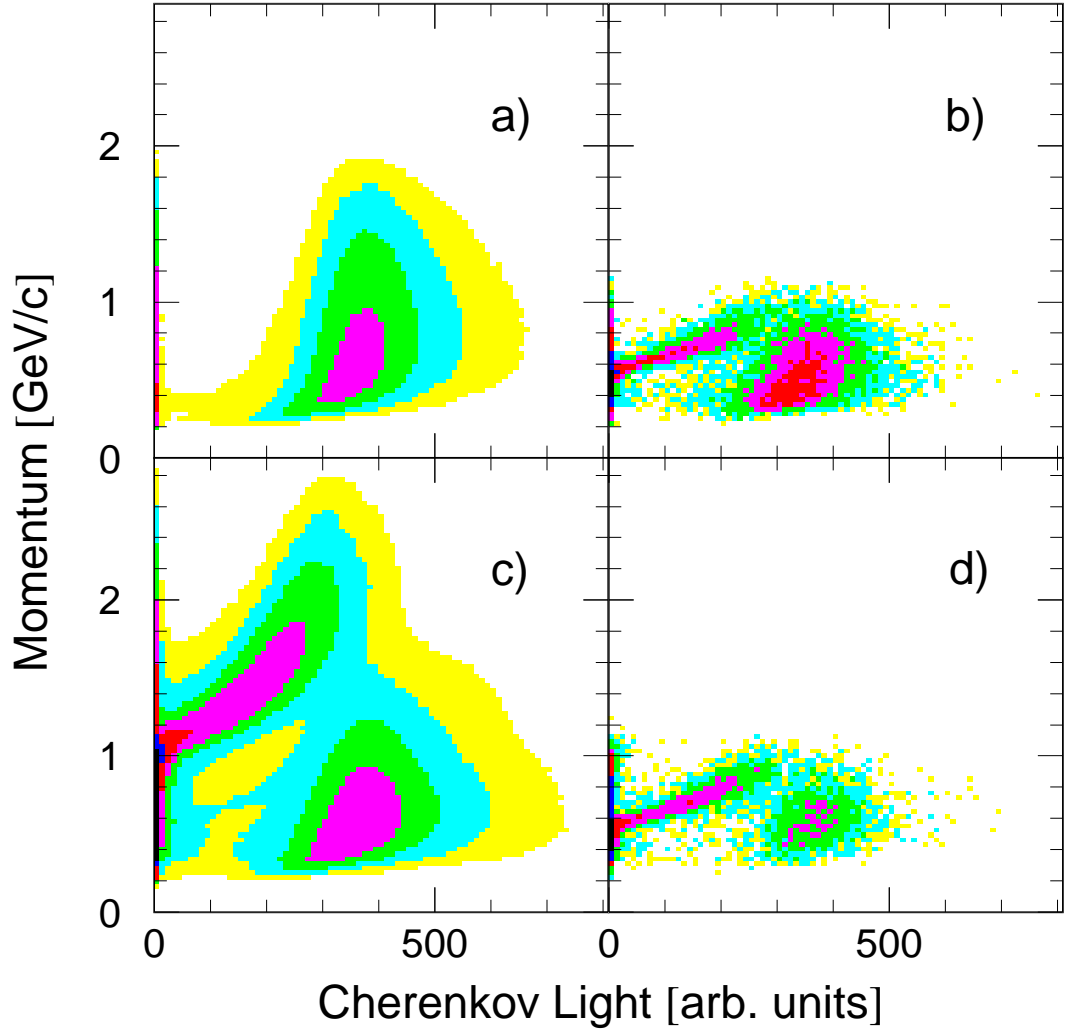}
  \end{center}
  \caption
  {Plots of the \v Cerenkov light output versus particle momentum for
    negatively(a and b) and positively(c and d) charged particles. The figures
    on the left are inclusive distributions and the figures on the right are
    the same distributions after requiring that the other \v Cerenkov
    amplitude is consistent with kaon identification and that the four
    observed particle momenta are consistent with 4-momentum conservation for
    the event hypothesis $ppK^+K^-$.  }
  \label{Fig:cerenkov}
\end{figure}

    The inclusive correlation of \v Cerenkov amplitude versus momentum is
    shown in Figures~\ref{Fig:cerenkov}a and \ref{Fig:cerenkov}c for
    negatively and positively curved tracks, respectively. These figures have
    a logarithmic intensity scale with a factor 2.4 between intensity
    contours.  In Figure~\ref{Fig:cerenkov}a there is a clear band from the
    $\pi^-$ mesons. The $K^-$ mesons are not visible in this inclusive
    distribution because of the $\sim10^5$ larger pion yield.  On the other
    hand, there is already a weak indication of the $K^+$ mesons in
    Figure~\ref{Fig:cerenkov}c in addition to the prominent $\pi^+$ meson and
    proton lines. The relative yield of a given kaon species can be enhanced
    by imposing two requirements on the data set: The first is that the
    oppositely charged meson is consistent with being a kaon, and the second
    is 4-momentum conservation for the given event hypothesis (i.e.
    $pp\rightarrow ppK^+K^-$).  The remaining yield after imposing these
    restrictions is shown in Figures~\ref{Fig:cerenkov}b and
    \ref{Fig:cerenkov}d, where signals from the $K^+$ and $K^-$ mesons are now
    clearly visible.

    \subsubsection{Event Reconstruction}
    The data sample analyzed here with four charged particles contained events
    from many different types of reactions.  The individual reactions were
    identified by a combination of particle identification of the charged
    particles in the final state and kinematic constraints on the measured
    particles.  The charged particles in the final state, predominantly
    protons, pions, and kaons, were identified with the water \v Cerenkov
    detectors as described above.  However, since the same final state could be
    produced via several reaction channels, kinematic conditions were
    required to discriminate among the particular reaction channels.
    
    All reaction channels studied here contained at most one unmeasured
    particle in the final state, thus the measured four track events were
    kinematically complete. As a result 4-momentum conservation was imposed
    as a powerful constraint to select events of a given hypothesis.  For this
    the invariant mass and missing mass of various particle configurations
    were calculated:  The invariant mass ($M_{inv}$) is the total energy in
    the reference frame of a given number of observed particles. $M_{inv}$ can
    be calculated from the total energies ($E$) and momenta ($\vec p$) of the
    $n$ individual particles using the following formula:
\begin{equation}
\left( M_{inv} \right)^2 = \left(\sum_{i=1}^n E_i \right)^2 - 
                           \left(\sum_{i=1}^n \vec p_i \right)^2.
\end{equation}
The missing mass analysis was used to determine the rest mass of an unobserved
recoil particle.  The missing mass ($M_{miss}$) is given by the following
formula:
\begin{equation}
\left( M_{miss} \right)^2 = \left(E_{beam} + M_p - \sum_{i=1}^m E_i \right)^2
                       - \left(\vec p_{beam} - \sum_{i=1}^m \vec p_i \right)^2
\end{equation}
where the index $i$ runs over all $m$ observed particles.  The specific
application of these constraints is discussed in the following sections.

\begin{figure}
  \begin{center}
    \epsfig{file=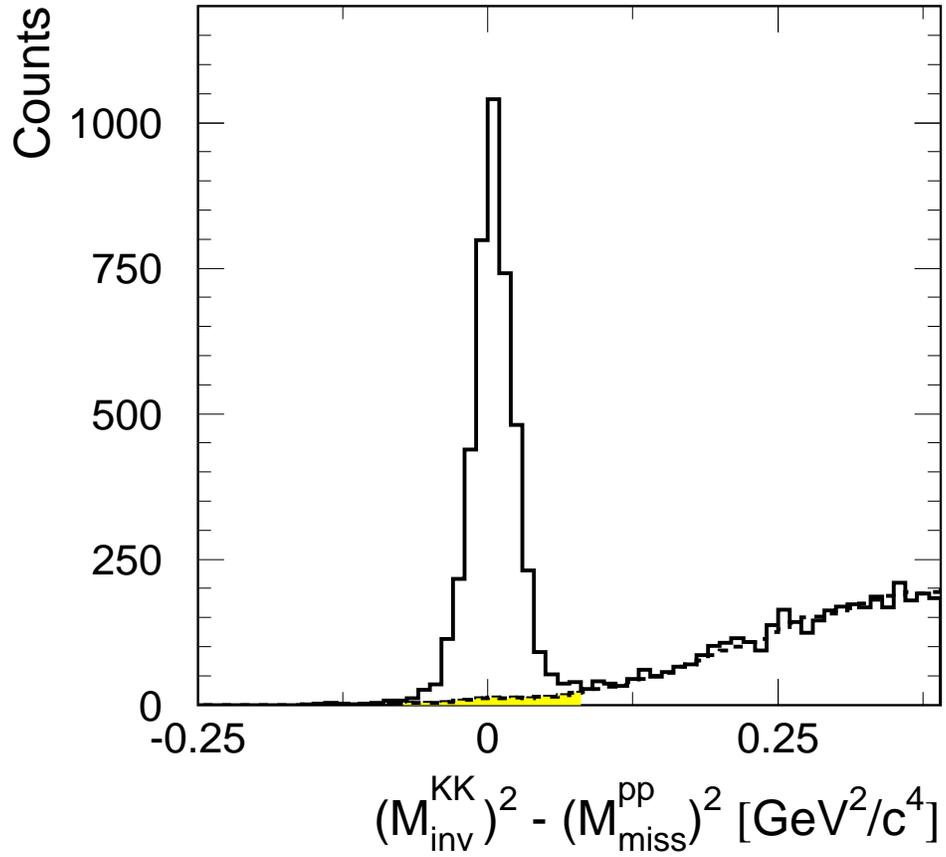,width=0.75\linewidth}
  \end{center}
  \caption{Distribution of $(M_{inv}^{KK})^2-(M_{miss}^{pp})^2$ for the event
    hypothesis $pp\rightarrow ppK^+K^-$. The dashed histogram is an estimate
    of the background by scaling the inclusive distribution without applying
    kaon \v Cerenkov requirements.}
  \label{fig:miss-inv}
\end{figure}

The results presented here on the production of the $\phi$ meson were
determined by selecting events of the type $pp\rightarrow ppK^+K^-$. For this
event hypothesis, the proton-proton missing mass ($M_{miss}^{pp}$) must equal
the invariant mass of the kaon pair ($M_{inv}^{KK}$).  Since all four
particles in the final state have been measured, the event reconstruction has
been performed requiring that the four particle missing mass
($M_{miss}^{ppKK}$) be equal to zero.  The distribution of
$(M_{inv}^{KK})^2 - (M_{miss}^{pp})^2$ is plotted in Figure~\ref{fig:miss-inv}
including kaon identification (solid histogram) based on the \v Cerenkov
detectors.  The peak at $(M_{inv}^{KK})^2-(M_{miss}^{pp})^2\approx 0$ results
from events consistent with the $pp\rightarrow ppK^+K^-$ hypothesis, and is
superimposed on a background resulting from imperfect $\pi-K$ separation in
the \v Cerenkov detectors in a small fraction of events of the type
$pp\rightarrow pK^+\Lambda \rightarrow ppK^+\pi^-$ or $pp \rightarrow
pp\pi^+\pi^-X$.  The dashed histogram is an estimate of the background by
scaling the inclusive distribution without kaon \v Cerenkov requirements by a
factor 0.002 in order to match the data in Figure~\ref{fig:miss-inv} above
$0.15~$GeV$^2/$c$^4$.  The background from non $ppK^+K^-$ final states
comprises only about 2.2\% of the yield in the range $|(M_{inv}^{KK})^2 -
(M_{miss}^{pp})^2| < 0.09$ GeV$^2$/c$^4$, which marks the range where events
were accepted for the further analysis.

\begin{figure}
  \begin{center}
    \epsfig{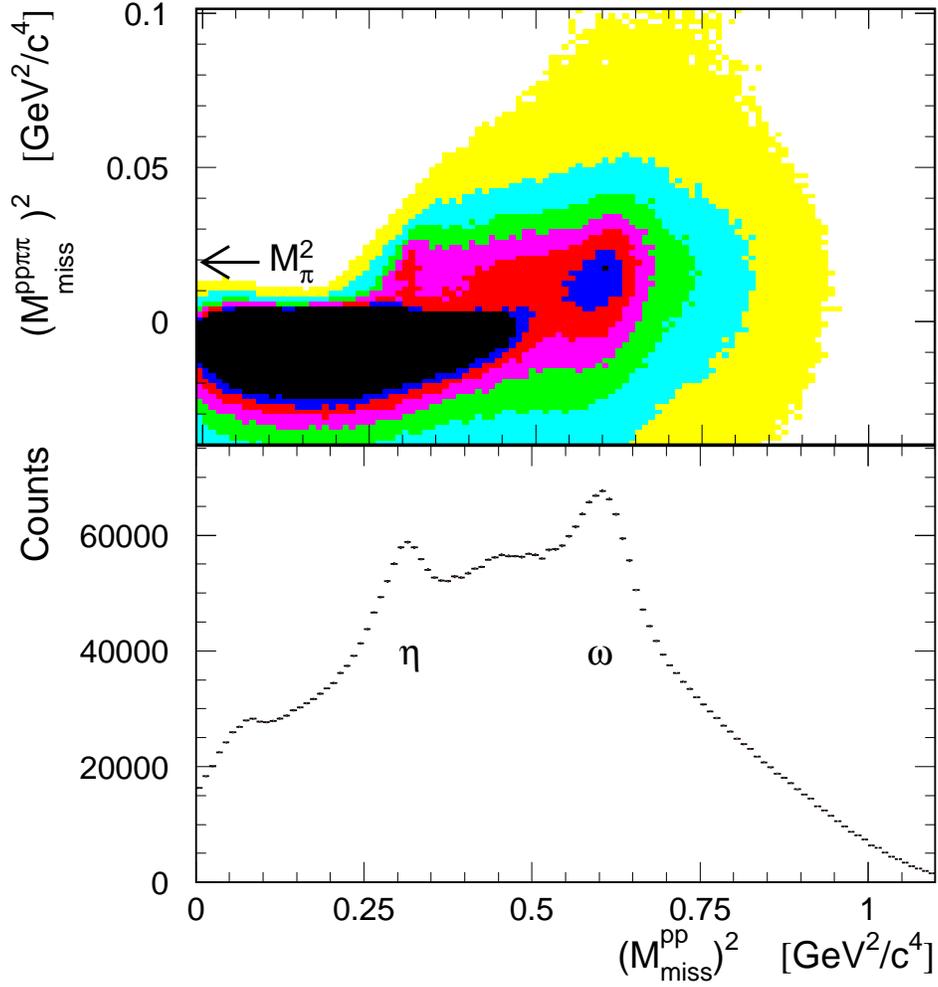}
  \end{center}
  \caption{Distribution of $(M_{miss}^{pp\pi^+\pi^-})^2$ versus
           $(M_{miss}^{pp})^2$ (upper frame). The lower frame shows
           prominent peaks from the $\eta$ and $\omega$ mesons in  
           the projection onto the $(M_{miss}^{pp})^2$ axis after
           requiring $(M_{miss}^{pp\pi^+\pi^-})^2 \approx (M_{\pi})^2$.}
  \label{fig:om_eta}
\end{figure}

The exclusive reaction channel $pp\to pp\omega$ was measured via the $\omega
\to \pi^+\pi^-\pi^0$ decay and was selected by a missing mass analysis since
the $\pi^0$ decays primarily into two photons
($\Gamma_{\gamma\gamma}/\Gamma_{\mbox{tot}}=0.988$) which are not detected in
the spectrometer.  Events of the type $pp\pi^+\pi^- \pi^0$ were selected by
first requiring that the \v Cerenkov amplitudes associated with the four
charged tracks be consistent with the hypothesis $pp\pi^+\pi^-$.  Furthermore,
background from
four body reactions of the type $pp\pi^+\pi^-$ with $M_{miss}^{pp}\sim
M_\omega$ is partially suppressed by requiring the invariant mass of the pion
pair to fulfill $(M_{inv}^{\pi^+\pi^-})^2 < (M_\omega - M_{\pi^0})^2$.  After
these requirements, the distribution of the four particle missing mass squared
$(M_{miss}^{pp\pi^+\pi^-})^2$ versus the proton-proton missing mass
$(M_{miss}^{pp})^2$ is shown in Figure~\ref{fig:om_eta} with a linear scale
for the intensity contours. As seen in this figure, there is a very strong
signal for the $pp\rightarrow pp\omega\rightarrow pp\pi^+\pi^-\pi^0$ reaction
at $(M_{miss}^{pp\pi^+\pi^-})^2 \approx (M_{\pi^0})^2$ and $(M_{miss}^{pp})^2
\approx (M_\omega)^2$. The projection of this distribution onto the
$(M_{miss}^{pp})^2$ axis is shown in the lower frame of
Figure~\ref{fig:om_eta} with the additional requirement that
$(M_{miss}^{pp\pi^+\pi^-})^2\approx (M_{\pi^0})^2$.  In this figure, there are
clear peaks from the $\pi^+\pi^-\pi^0$ decay of the $\eta$ and $\omega$
mesons.  The structure at very low $(M_{miss}^{pp})^2$ is due to contamination
of four body events of the type $pp\pi^+\pi^-$.

  \subsection{Acceptance Correction}            \label{ssec:accpt}
  The relative acceptance of the apparatus for the $pp\eta$, $pp\omega$ and
  $ppK^+K^-$ production channels has been evaluated by means of Monte Carlo
  simulations, which after digitization, were processed through the same
  analysis chain as the measured data. The relative acceptance could be
  determined independent of the actual phase space distribution of the
  particles in the final state, because the following two requirements are
  fulfilled with the DISTO spectrometer: (i) the detector acceptance was
  determined as a function of all relevant degrees of freedom in the final
  state, and (ii) after accounting for the azimuthal and reflection
  symmetries, the detector acceptance was non-zero over the full kinematically
  allowed region.
  
  In order to determine the detector acceptance, the relevant degrees of
  freedom were each subdivided into a specific number of bins, thus defining a
  multi-dimensional matrix.  Next, the number of both the generated and the
  reconstructed events from the simulations were stored in separate copies of
  this matrix.  Finally, the bin-by-bin ratio of the number of generated to
  the number of reconstructed events provided the efficiency correction
  matrix.
  
  When analyzing the data the acceptance correction procedure was performed on
  an event-by-event basis by first determining the phase space bin of the
  particular event and then incrementing the distributions by the
  corresponding acceptance correction factor from the acceptance correction
  matrix described below. These correction factors include the appropriate
  partial widths and the simulations account for the lifetime and different
  decay modes of all unstable particles involved.

    \subsubsection{$p p K^+ K^-$ final state}
    Although there are in general 16 degrees of freedom for 4 particles in the
    final state, numerous constraints exist which significantly lower the
    total number of independent degrees of freedom.  For instance, particle
    identification and 4-momentum conservation reduce the number of degrees of
    freedom to 8 for the $p p K^+ K^-$ final state. The remaining degrees of
    freedom can be parameterized in numerous ways, with the only restriction
    that the variables chosen must span the space of allowed possibilities.
    
    Without loss of generality the eight kinematic degrees of freedom for the
    $ppK^+K^-$ final state can be parameterized as if the reaction proceeded
    in two steps, i.e.  $pp\rightarrow p_1 p_2 X \rightarrow p_1 p_2 K^+K^-$.
    In this case, five variables can be used to describe the three body system
    $ppX$, and the remaining three degrees of freedom are then required to
    uniquely determine the decay of the intermediate state $X$.  This choice
    of kinematic variables provides a complete basis and does not require that
    the reaction actually proceeds via the intermediate state $X$.  The $ppX$
    system is parameterized by the two invariant mass combinations
    $(M_{inv}^{p_1X})^2$ and $(M_{inv}^{p_2X})^2$ (i.e. Dalitz plot variables
    of the $ppX$ system), and three Euler angles to describe the orientation
    of the $ppX$ decay plane in the center of mass reference frame (CM): the
    polar and azimuthal angles of the $X$ intermediate state $\Theta_{CM}^X$,
    $\phi_{CM}^X$, and a rotation of the $ppX$ decay plane around the
    direction of $X$, $\psi_{CM}^{pp}$.  Furthermore, three variables are
    required to describe the decay of $X$: the mass $M_X=M_{inv}^{KK}$, and
    the emission angles of one of the kaons in the $X$ reference frame
    $\Theta_X^K$ and $\phi_X^K$ where these angles are measured relative to
    the beam direction\footnote{The following conventions are used to label
      the various angles that appear in this report: $\Theta$ corresponds to
      polar angles measured with respect to the beam axis, and $\Psi$
      corresponds to polar angles measured with respect to any other
      quantization axis, which will be explicitly stated.  Azimuthal angles
      are labeled by $\phi$ and $\psi$, respectively.  The subscript denotes
      the reference frame in which the angle in measured, and the superscript
      denotes the particle being measured.}.  The beam direction is chosen as
    the reference direction for the $X$ decay to facilitate comparison in
    Sec.~\ref{ssec:k+k-} with simple expectations for $\phi$-meson decay when
    it is produced near threshold in $pp$ reactions.

    The differential cross section for $X$
    production must be isotropic as a function of the azimuthal angle
    $\phi_{CM}^X$.  Therefore, in the simulations $\phi_{CM}^X$ has been
    integrated with an isotropic distribution to determine the detector
    acceptance as a function of the remaining seven degrees of freedom.
    
    Due to the high dimensionality of this problem, even a small number of
    bins per kinematic variable result in an acceptance correction matrix that
    is too large to calculate with the available computational facilities.
    For a subdivision of each degree of freedom into $N= 10$ bins, and
    an average of $M= 20000$ events simulated per bin of the acceptance
    matrix, a total of $M\times N^7\approx 2\times 10^{11}$ would have to be
    simulated.  Thus, in order to explore the dependence of the detector
    acceptance on the remaining seven kinematic variables, two subsets of the
    general acceptance correction matrix (Matrix A and Matrix B) have been
    generated by integrating over several of the kinematic variables as
    discussed below and summarized in Table~\ref{tab:KinVarPhiEtaOm}.
    
    The simulations for Matrix A contain two simplifications in order to make
    the calculations tractable. The first simplification is the requirement
    that the mass of the intermediate state $X$ be equal to the $\phi$ meson
    mass.  Of course this requirement restricts the use of Matrix A to the
    $pp\phi$ reaction only; however, little systematic error is introduced
    because the shape of the mass distribution of the $\phi$ meson is well
    known and the detector acceptance varies little over it. Furthermore, an
    isotropic angular distribution of the differential cross-section for
    $\phi$ meson production versus $\Theta_{CM}^\phi$ has been assumed for
    Matrix A. This assumption is consistent with the observation that this
    angular distribution turns out to be isotropic when Matrix B (see below),
    which includes explicitly the $\Theta_{CM}^X$ dependence, is used for the
    acceptance correction.  Therefore, in the simulations to calculate Matrix
    A, the intermediate state $X$ was given the $\phi$ meson mass and the
    angular variables $\Theta_{CM}^X$ and $\phi_{CM}^X$ were integrated over
    with isotropic distributions when determining the acceptance as a function
    of the remaining five variables.
        
    In order to overcome the restrictions on the usage of the acceptance
    correction Matrix~A mentioned above, the available degrees of freedom in
    the $ppK^+K^-$ final state were subdivided differently, as listed in
    Table~\ref{tab:KinVarPhiEtaOm}, resulting in the acceptance correction
    Matrix B.  Here the variables $M_{inv}^{KK}$ and $\Theta_{CM}^X$ were
    explicitly included in Matrix B.  However, an isotropic distribution for
    the decay $X \to K^+K^-$ was used in order to reduce the dimensionality of
    the matrix to a solvable level.  The validity of using an isotropic decay
    of $X \to K^+K^-$ can be judged by the data presented in
    Sec.~\ref{ssec:k+k-} using Matrix A.  Furthermore, the variables $\left(
      M_{inv}^{p_1X}\right)^2$ and $\left( M_{inv}^{p_2X}\right)^2$ were
    integrated over with a weighting according to three body phase space.
    This was motivated by the observation that the physical distributions show
    only small deviations from isotropy and double checked to give consistent
    results with a different generator for the $ppX$ system.

    \subsubsection{$p p \pi^+ \pi^- \pi^0$ final state}
    For the measurements of the $\eta$ and $\omega$ meson production there are
    5 particles in the final state of each channel observed.  The 20 degrees
    of freedom associated with 5 particles are reduced to 12 because of
    4-momentum conservation and four particles have been identified.  Two
    further degrees of freedom are eliminated by requiring $M_{miss}^{pp} =
    M_{\eta, \omega}$, respectively, and $M_{miss}^{pp\pi^+\pi^-}=M_{\pi^0}$.
    Of the remaining 10 degrees of freedom, five are related to the $ppX$
    system where $X=\eta ,\omega$ and five to the corresponding decay $X \to
    \pi^+ \pi^- \pi^0$.
    
    The $ppX$ system is parameterized by the Dalitz plot variables
    $(M_{inv}^{p_1X})^2$ and $(M_{inv}^{p_2X})^2$, as well as three Euler
    angles to describe the orientation of the $ppX$ decay plane:
    $\Theta_{CM}^X$, $\phi_{CM}^X$, and $\psi_{CM}^{pp}$.  As discussed above,
    the differential cross section is isotropic in $\phi_{CM}^X$, thus there
    are in effect 4 independent degrees of freedom plus those associated with
    the decay of the meson $X$.  The matrix element associated with the
    $\omega\rightarrow\pi^+\pi^-\pi^0$ decay was taken from~\cite{Lic94} and
    verified to be consistent with the data from~\cite{Ste62,AST93}, thus
    allowing the corresponding variables to be integrated over. The matrix
    element for the $\eta$ decay was taken from \cite{Abe98,Ams95b,Lay73}.
    Furthermore, we have assumed an isotropic orientation of the $\omega$
    decay plane, which was verified to be consistent with the data.  Thus,
    four dimensional efficiency matrices were calculated for the $\eta$ and
    $\omega$ production reactions, as summarized in
    Table~\ref{tab:KinVarPhiEtaOm}.
    
    Although all bins that are kinematically allowed have a finite acceptance,
    there are some phase space bins with very low acceptance.  These bins are
    associated with backward emission of the $\eta$ or $\omega$ meson (in the
    CM frame). Because the initial system involves two identical particles,
    the physical distribution must have a symmetry about $\Theta_{CM}^X =
    90^\circ$.  Thus, in order to reduce the systematic error associated with
    the very large acceptance corrections at backward angles, all integrated
    cross section results for the $\eta$ and $\omega$ meson production were
    calculated for the forward hemisphere only (and then multiplied by two).
    The differential cross sections however could be extended to backward
    angles, up to the limits of where the acceptance correction method
    remained valid, thereby providing an additional check of the calculations.

  \subsection{Absolute Normalization}           \label{ssec:absnorm}
  
\begin{figure}
  \begin{center}
    \epsfig{file=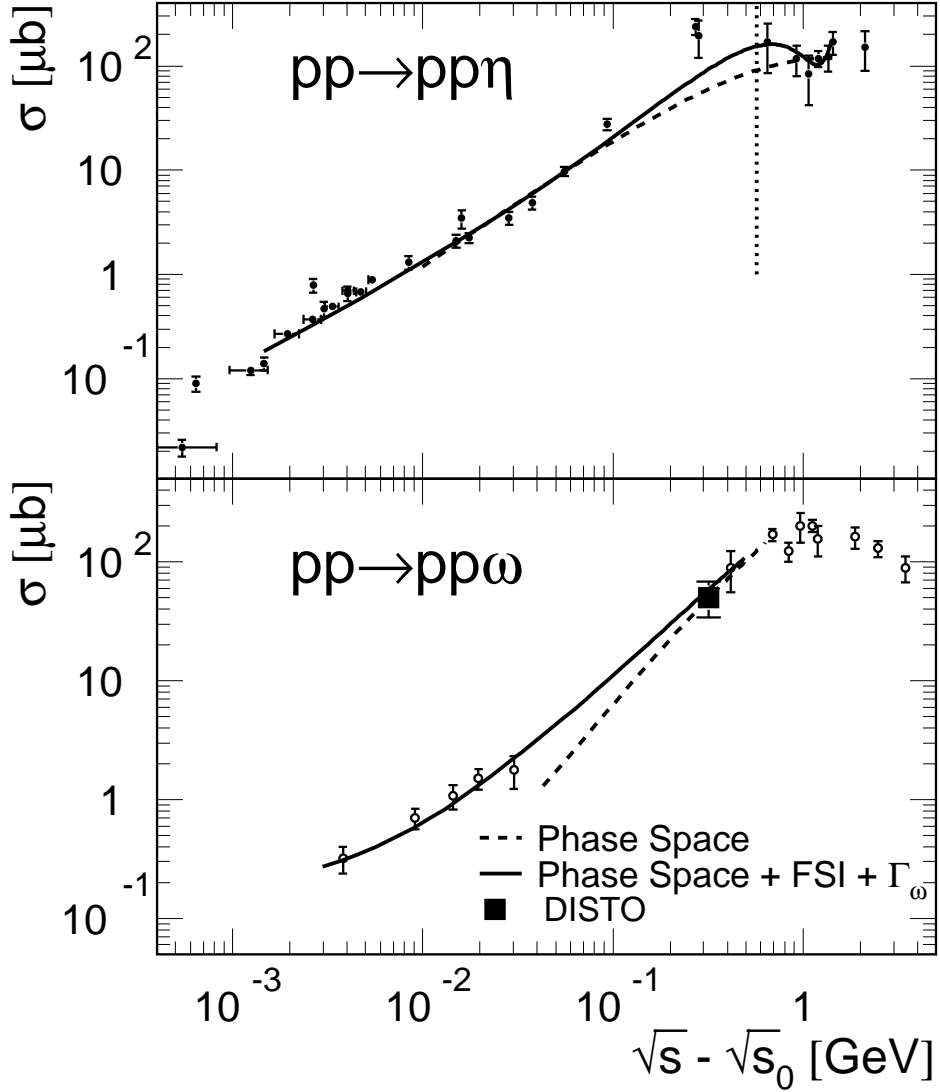,width=0.75\linewidth}
  \end{center}
  \caption{Exclusive cross section as a function of the total
    CM energy above threshold for the reaction $pp\rightarrow pp\eta$ (upper
    frame) and of the reaction $pp\rightarrow pp\omega$ (lower frame).  The
    data points are referenced in the text.  The solid and dashed curves in
    the upper frame are parameterizations of the data in order to estimate the
    $\eta$ meson production cross section at the energy in this experiment,
    which is marked by the vertical dotted line.  The filled square in the
    lower frame is the result of this experiment.  In the lower frame the
    dashed curve shows the CM energy dependence of the three-body ($pp\omega$)
    phase space volume. The solid curve additionally includes the finite width
    of the $\omega$ meson and the proton-proton final state interaction
    (FSI).}
  \label{fig:eta_omega_xc}
\end{figure}

  The absolute cross section normalization was determined by measuring the
  yield of a given channel relative to that of a {\em simultaneously} measured
  channel with known cross section.  This method to determine the absolute
  normalization was chosen because it reduced the large systematic
  uncertainty associated with the absolute calibrations of both beam intensity
  and trigger efficiency.  For this work the reference channel was the
  reaction $pp\rightarrow pp\eta$, for which a large amount of existing data
  \cite{Smy00,Cal96,Ber93,Chi94,Pic62,Ale67,Bod68,Col67,Cas68,Col70,Yek70,Alm68}
  are summarized in the upper frame of Figure~\ref{fig:eta_omega_xc}.
  
  In order to provide the absolute cross section calibration, the existing
  published data were interpolated to estimate the $\eta$ production
  cross-section at the beam momentum of the present measurement.  The solid
  and dashed curves presented in the upper frame of
  Figure~\ref{fig:eta_omega_xc} correspond to two different parameterizations
  of the measured cross section values.  The solid curve corresponds to a
  polynomial of sixth order and the dashed curve corresponds to a
  parameterization of the following form:
\begin{equation}
\sigma_{pp\rightarrow pp\eta} =
a \left(\frac{s}{s_0}-1\right)^b \left(\frac{s_0}{s}\right)^c.
\end{equation}
  where $\sqrt{s_0}=(2 M_p + M_\eta)$ is the CM energy at the $\eta$ meson
  production threshold, and $a, b$ and $c$ are free parameters.  The vertical
  dotted line marks the available energy ($Q = 0.554$~GeV) of this measurement.
  
  Both parameterizations describe the existing $\eta$ total cross section data
  well with the exception of the measurement at $p_{beam}=2.8$~GeV/c by
  E.~Pickup et al.~\cite{Pic62}, which is underestimated. In the upper frame
  of Figure~\ref{fig:eta_omega_xc}, both values cited in Ref.~\cite{Pic62},
  for identification via the $\eta \rightarrow \pi^+\pi^-\pi^0$ and $\eta
  \rightarrow neutrals$ decay channels are plotted at the corresponding CM
  energy above threshold $Q = 0.275$~GeV.  This discrepancy has been neglected
  since that measurement~\cite{Pic62} is subject to a large systematic error
  associated with a quite substantial background subtraction.  The average of
  these two interpolations at $Q = 0.554$~GeV is our estimate of $135\pm
  35~\mu$b for the total cross section of the reaction $pp\rightarrow pp\eta$
  at our beam momentum. The systematic error from the absolute normalization
  ($\pm 26\%$) is determined from the range of the parameterizations and is
  similar to or smaller than the systematic error from the combination of all
  other sources (i.e. 32\% for the $\phi/\eta$ ratio, see
  Table~\ref{tab:uncert}). Furthermore, our estimate is in good agreement with
  one boson exchange model calculations by Vetter et al.~\cite{Vet91}, who
  predict $\approx 120~\mu$b at this energy.

  \subsection{Systematic Errors}                \label{ssec:syserr}
  Due to the large amount of data collected, the statistical errors are
  relatively small, and the experimental error is dominated by systematic
  uncertainty and systematic bias.  This section summarizes the effects
  studied to estimate the magnitude of the systematic uncertainty, as well as
  systematic deficiencies of the acceptance correction method and trigger
  bias.
  
  Systematic uncertainty of the results quoted here have been studied in
  detail for the following effects:
  \begin{itemize}
  \item For the acceptance corrections, the statistical uncertainty of the
    simulations has been studied by comparing the reconstructed particle
    ratios using acceptance correction matrices based on different subsets of
    the simulated data. Furthermore, effects due to the finite binning of the
    kinematic variables, as well as the finite detector resolution have
    been considered. Moreover, an additional cross check of the acceptance
    corrections was provided by the observation that the total $\phi$ meson
    yield is identical (within the error range) when determined using both
    acceptance correction Matrices A and B.
  \item A significant source of uncertainty is related to the meson yield
    determination (back- ground subtraction and peak shape parameterization).
    This effect was studied by varying the parameterization of the line shapes
    and background, as well as by varying the fit ranges.
  \item The four track trigger used to record the data presented here had a
    slightly different efficiency for $pp\pi^+\pi^-$ and $ppK^+K^-$ events.
    Although this effect has been corrected for, as discussed below, effects
    such as noise and cross-talk in the multi-anode photomultipliers used for
    readout of the scintillating fiber detectors introduce an uncertainty to
    the magnitude of this correction, which we denote as ``trigger bias".
  \item Similarly, the tracking efficiency of the data analysis procedure
    may vary slightly between measured and simulated data, primarily due to
    uncertainty of the actual wire chamber efficiency. In particular, the
    single track efficiency and correlated efficiency losses due to small
    spatial separation between tracks have been studied in detail.
  \item Since the data presented here were collected over an extended period
    of time, residual efficiency losses due to long term drifts in the
    calibration of the electronics may remain.  This effect was examined by
    comparing the reconstructed particle ratios in temporally separated
    subsets of the data.
  \item Finally, uncertainty arising from the particle identification in the
    \v Cerenkov detectors and the kinematic conditions has been estimated by
    comparing the efficiency loss due to each individual restriction between
    the simulated and the measured data.
  \end{itemize}    
  These results are summarized in Table~\ref{tab:uncert}.  The total
  systematic uncertainty quoted is determined by quadratically adding the
  individual terms applicable for a given measured cross section ratio.
    
  In addition to the systematic uncertainty, several effects lead to a
  systematic bias of the calculated cross section ratios. To estimate the
  magnitude of the systematic bias we have examined the following effects.
  \begin{itemize}
  \item The liquid hydrogen target was contained in a vessel in which
    background reactions were produced. These ``non-target'' events could be
    well separated by the vertex location in those events with complete
    scintillating fiber information. Since this separation was not possible in
    the remaining $\approx$ 50\% of the events, the relative contamination
    determined in those events with full fiber information was subtracted from
    the complete data sample.
  \item The data collection trigger required that three of the four charged
    particles must be observed in the scintillating fiber detectors.  Since
    the fiber detectors had a different response for minimum ionizing pions
    compared to kaons, which were slower, the trigger efficiency was higher for
    $ppK^+K^-$ events than for $pp\pi^+\pi-$ events. The variation of the
    fiber efficiency has been studied as a function of particle velocity. The
    magnitude of the trigger bias was thus determined by combining the average
    kaon and pion efficiencies with the ``three of four'' trigger condition
    and an estimate of the cross talk in the fiber detectors.
  \item Finally, in the data analysis, conditions were placed on the \v
    Cerenkov amplitude of the individual tracks, as well as on several
    kinematical quantities.  The efficiency with which a particular event type
    is accepted by all these conditions is strongly related to the detector
    resolution, which may be imperfectly modeled in the simulations, leading
    to a bias in the efficiency correction matrices.  The magnitude of this
    effect was examined for each meson production reaction by comparing the
    acceptance loss from each individual selection criterion in the data to
    the simulations.
  \end{itemize}
  The individual contributions to the systematic bias are summarized in
  Table~\ref{tab:bias} as the factor by which they modify the given particle
  cross section ratio.  Based on these considerations, the reconstructed $
  \phi/\omega$, $\phi/\eta$, and $\omega/\eta$ cross section ratios have been
  multiplied by the factors 0.97, 0.98, and 1.02, respectively, to account for
  the systematic bias. The correction factor for the total $K^+K^-/\eta$ cross
  section ratio is the same as for the $\phi/\eta$ ratio.

\section{Results}                               \label{sec:results}
After full acceptance corrections, the relative yields for exclusive
$\eta,\omega$, and $\phi$ meson production in proton-proton reactions at
3.67~GeV/c have been determined.  These results are summarized in
Table~\ref{tab:ratios}, including the statistical and systematic error, and
have been combined with the absolute $\eta$ meson production cross section
(see Sect.~\ref{ssec:absnorm}) to determine the absolute cross section results
presented below.

  \subsection{$\omega$ Meson Production}        \label{ssec:etaOmega}
  
  Using the $\eta$ meson yield as the absolute normalization, as discussed
  above, the total cross section for the reaction $pp \to pp\omega$ is
  determined to be $(50\pm3\ ^{+18}_{-16})~\mu$b with the statistical and
  systematical error, respectively.  This value is plotted as the solid point
  in the lower frame of Figure~\ref{fig:eta_omega_xc}.  The other data points
  are from references \cite{Bod68,Col67,Cas68,Col70,Yek70,Alm68,Hib99}.
  
  The dashed curve shows the energy dependence according to three-body
  ($pp\omega$) phase space and has been normalized to the point
  directly above this measurement (i.e.  $\sqrt{s}-\sqrt{s_0}=0.41$~GeV).  The
  solid curve shows the expected behavior when taking into account the finite
  width of the $\omega$ meson and the proton-proton final state interaction
  (FSI)~\cite{Sib99a,Sib99b}.  Although the energy of this experiment is
  sufficiently high that pure S wave production is not expected, a smooth
  variation of the cross section with $\sqrt{s}$ is expected since there are
  no known baryonic resonances with significant branching ratios to $p\omega$.
  Thus, the good agreement of the total cross section for $\omega$ meson
  production determined here with the curves in the figure, is a strong
  indication that the absolute normalization used here does not introduce an
  error larger than the quoted systematic error.  Furthermore, our result is
  in good agreement with the value $(45\pm 7)~\mu$b determined in~\cite{Sib99b}
  by an interpolation between the existing data.

\begin{figure}
  \begin{center}
    \epsfig{file=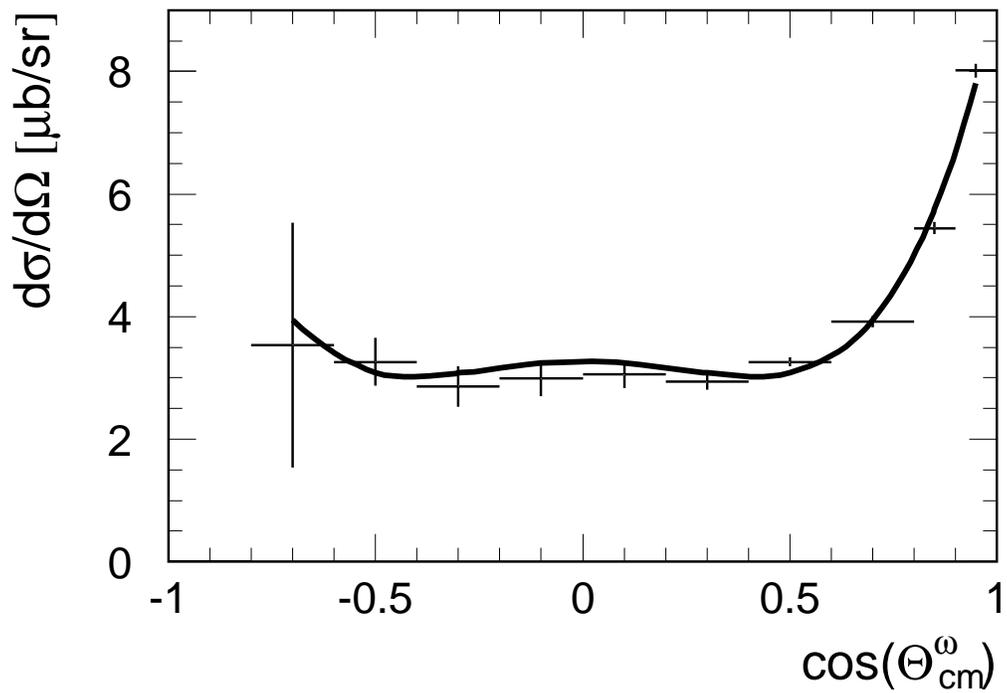,width=0.8\linewidth}
  \end{center}
  \caption{ Differential 
    cross-section (in the CM frame) for the $pp\to pp\omega$ reaction as a
    function of $\cos \Theta^\omega_{CM}$.  The solid curve is a fit to the
    data with the sum of the first three even Legendre polynomials.}
  \label{fig:OmAngDist}
\end{figure}

  The differential cross-section for $\omega$ meson production has been
  plotted versus $\cos \Theta^{\omega}_{CM}$ in Figure~\ref{fig:OmAngDist}.
  This distribution has been fit with the sum of the first three even Legendre
  polynomials $P_i$. The best fit has been obtained with the expression below
  and shown as the solid curve in the figure.
  \begin{equation}
  \frac{d\sigma}{d\Omega} = (4.0\pm0.1)\cdot P_0 +
                           ( 3.1\pm0.2)\cdot P_2 +
                           ( 2.0\pm0.2)\cdot P_4  \hspace{4mm}(\mu\mbox{b/sr})
  \end{equation}
  Due to the symmetry of the incoming channel, this angular distribution must
  have a reflection symmetry about $\cos \Theta_{CM}^\omega = 0$.  Despite the
  strong variation of the detector acceptance with $\Theta_{CM}^{\omega}$,
  this symmetry is observed in the data, providing a very important
  confirmation of the validity of the detector acceptance correction method.
  The deviations from isotropy indicate that partial waves up to $L=2$
  for the $\omega$ meson relative to the $pp$ system are involved in the
  production mechanism.  Further information such as from polarization degrees
  of freedom are however required in order to make a quantitative measure of
  the relative partial wave amplitudes.

  \subsection{$\phi$ Meson Production}          \label{ssec:k+k-}
\begin{figure}
  \begin{center}
    \epsfig{file=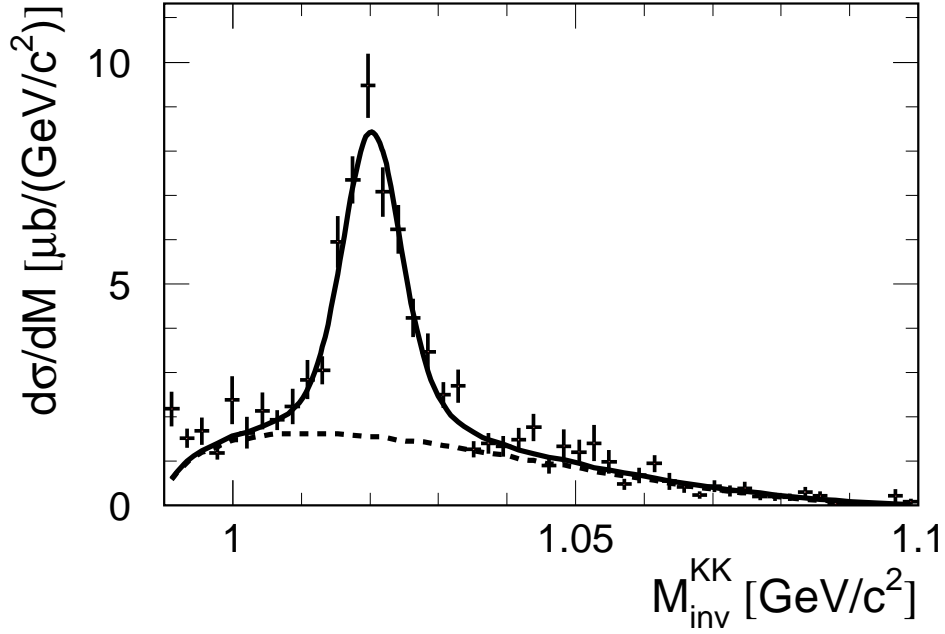,width=0.75\linewidth}
  \end{center}
  \caption{ Acceptance
    corrected $M_{inv}^{KK}$ distribution. The dashed curve is an estimate of
    the non-resonant contribution, and is based on the $M_{inv}^{KK}$
    dependence of four body phase space ($ppKK$). The solid curve is the sum of
    the non-resonant and the $\phi$ meson contributions.}
\label{fig:KKMatA}
\end{figure}

  The $M_{inv}^{KK}$ distribution is shown in Figure~\ref{fig:KKMatA} after
  application of the acceptance correction Matrix B (which explicitly includes
  $M_{inv}^{KK}$) and the absolute normalization discussed above.  This
  distribution has been analyzed to determine what fraction of the yield is
  due to non-resonant $K^+K^-$ production and what fraction has been produced
  via the $\phi$ resonance.  The dashed curve shows the estimated non-resonant
  contribution. The shape of this contribution is given by the $M_{inv}^{KK}$
  distribution for final states distributed according to four particle phase
  space ($ppK^+K^-$).  The shape of the resonant contribution is given by the
  natural line shape of the $\phi$ meson folded with a Gaussian to account for
  the detector resolution.  The width of this Gaussian is
  $\sigma=3.3\pm0.5$~MeV, in good agreement with the simulations
  ($\sigma=3.4\pm0.1$~MeV).  The total cross section for the $pp\to ppK^+K^-$
  reaction, as well as the resonant and non-resonant contributions are
  summarized in Table~\ref{tab:KKCross}.  In addition, the total cross section
  for $\phi$ meson production has been included after correcting for the
  branching ratio $(\Gamma_{K^+K^-}/\Gamma_{tot} = 0.491)$ and rounding to two
  significant digits.

  
  The $pp\phi$ final state can be defined by two angular momenta, $l_1$ is the
  orbital angular momentum of the two nucleons relative to each other and
  $l_2$ is the orbital momentum of the $\phi$ meson relative to the two
  nucleon CM system.  At threshold the $pp\phi$ final state has the orbital
  angular momenta $l_1 = l_2 = 0$.
  
\begin{figure}
  \begin{center}
    \epsfig{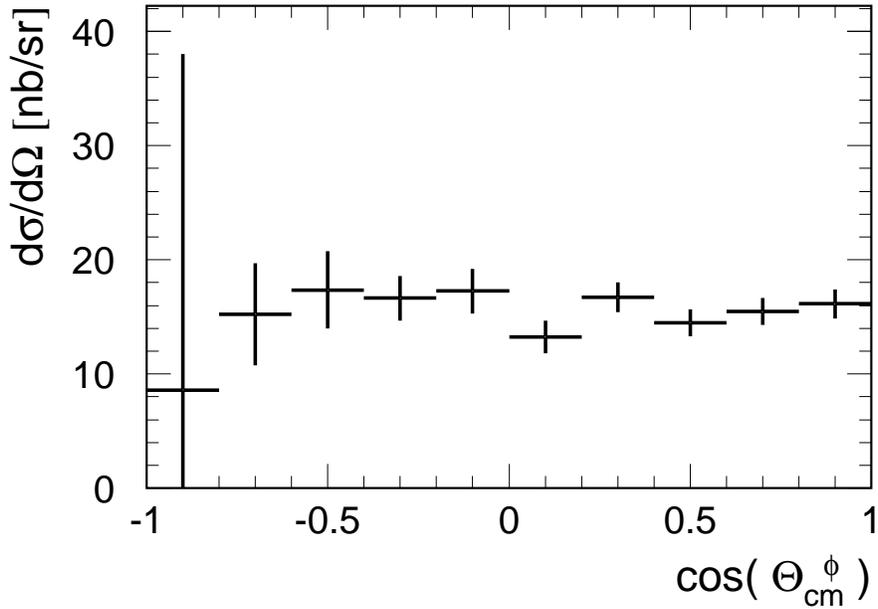}
  \end{center}
  \caption{Differential cross section for $\phi$ 
    meson production as a function of $\cos \Theta^\phi_{CM}$.}
  \label{fig:1080}
\end{figure}

  The differential cross-section for $\phi$ meson production is plotted versus
  $\cos \Theta_{CM}^\phi$ in Figure~\ref{fig:1080}.
  This distribution must be symmetric around $\cos \Theta_{CM}^\phi = 0$.  The
  fact that the observed distribution is indeed symmetric about $\cos
  \Theta_{CM}^\phi = 0$ (within the error bars), as it was already observed in
  Figure~\ref{fig:OmAngDist} for the angular distribution
  for the $\omega$ meson, is an additional consistency check for the validity
  of the acceptance corrections.  Furthermore, no significant deviations from
  isotropy are seen in these data, indicating that the $\phi$ is predominantly
  in a S wave state relative to the two protons (i.e. $l_2 = 0$).  This
  observation is in good agreement with the expectations of Rekalo et
  al.~\cite{Rek97} who claim $l_2=0$ up to a $\phi$ meson CM momentum of
  $p_\phi^* \le M_K c$ which corresponds to an available energy $Q\approx
  178$~MeV (in comparison, this measurement is at $Q=83$~MeV).
  
\begin{figure}
  \begin{center}
    \epsfig{file=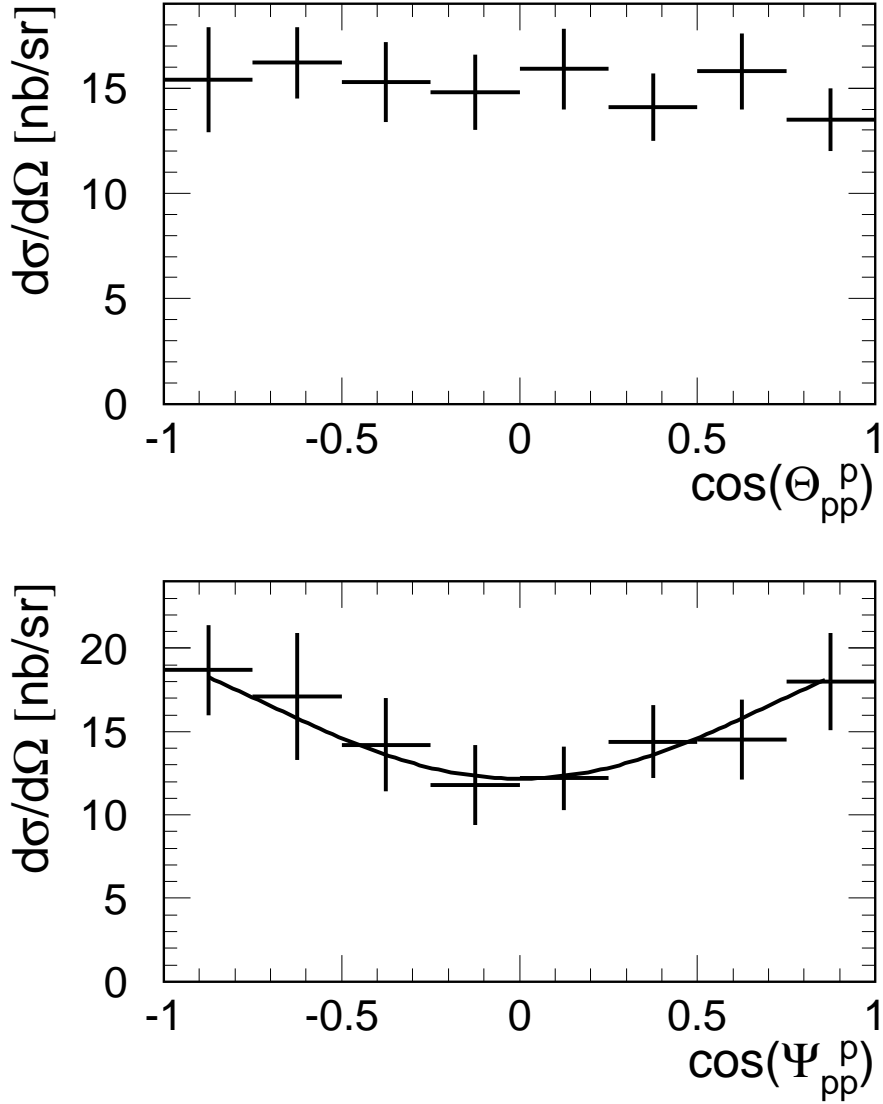,width=0.7\linewidth}
  \end{center}
  \caption{ (Upper frame) Differential cross section for 
    the $pp\to pp\phi$ reaction plotted as a function of $\cos \Theta^p_{pp}$,
    measured in the $pp$ reference frame relative to the beam direction.
    (Lower frame) Differential cross section for the $pp\to pp\phi$ reaction
    plotted as a function of $\cos \Psi^p_{pp}$, measured in the $pp$
    reference frame relative to the direction of the $\phi$ meson. The solid
    line is a fit to the data with the sum of the first three even Legendre
    polynomials.  }
  \label{fig:70}
\end{figure}

  Moreover, Rekalo suggests that the $pp$ system may be excited to higher
  partial waves at much smaller energies above threshold than needed to excite
  the $\phi$ meson relative to the protons (i.e. $l_1=1, l_2=0$ should occur
  at lower energies than $l_1=0, l_2=1$).  To test this expectation the
  proton-proton angular distribution has been evaluated for the events with
  $\phi$ meson production.  In the upper frame of Figure~\ref{fig:70}, the
  differential cross section is plotted as a function of the polar angle under
  which the protons are emitted relative to the beam, measured in the
  proton-proton reference frame.  Here, the observed angular distribution is
  consistent with being isotropic.  On the other hand, the proton-proton
  angular distribution exhibits a significant deviation from isotropy, when
  measured relative to the direction of the $\phi$ meson, as presented in
  the lower frame of Figure~\ref{fig:70}.  This distribution has been
  parameterized with the sum of the first three even Legendre polynomials as
  listed below and shown as the solid curve in the figure.
  \begin{equation}
  \frac{d\sigma}{d\Omega} = (15.0\pm0.9)\cdot P_0 +
                          ( 5.1\pm2.2)\cdot P_2 -
                          ( 0.7\pm3.6)\cdot P_4  \hspace{4mm} (\mbox{nb/sr})
  \end{equation}
  Odd Legendre polynomials have been omitted due to the reflection symmetry of
  the final $pp$ state about its CM motion direction, which is opposite the
  $\phi$ momentum in the overall CM frame.  These data are well described by
  using only the lowest two Legendre polynomials, as evident by the large
  uncertainty associated with the $P_4$ coefficient.  These results 
  indicate that partial waves up to $l_1=1$ are involved
  in the proton-proton exit channel.

  The momentum distribution of the particles in the final state is also
  related to the relative partial wave contributions in the $pp\phi$ system.
  Defining $p$ to be the momentum of a proton in the $pp$ reference frame and
  $q$ to be the CM momentum of the $\phi$ meson, the total cross section can
  be written as the sum of the individual partial wave
  contributions~\cite{Newton}:
\begin{equation}
\sigma \sim \sum_{l_1,l_2} \int  |M_{l_1,l_2}|^2 d\rho_{l_1l_2}.
\end{equation}
  Here $M_{l_1,l_2}$ is the matrix element for a given final state with
  angular momenta $l_1,l_2$, and $d\rho_{l_1l_2}$ is an element of the
  three body phase space which is given by the following formulae.
\begin{equation}
d\rho_{l_1l_2} \sim p^{2l_1+1}q^{2l_2+2}dq, 
\end{equation}
  where the proton momentum is given by
\begin{equation}
p = \sqrt{q^2_{max}-q^2} \times \sqrt{\frac{1}{4}+\frac{m_p}{2m_\phi}},
\end{equation}
  and the maximum CM momentum of the $\phi$ meson and the
available energy are
\begin{equation}
q_{max} = \sqrt{\frac{4m_pm_\phi Q}{m_\phi + 2m_p}},
\hspace{1.5cm}
Q= \sqrt{s} - \sqrt{s_{0}}.
  \end{equation}
  
\begin{figure}
  \begin{center}
    \epsfig{file=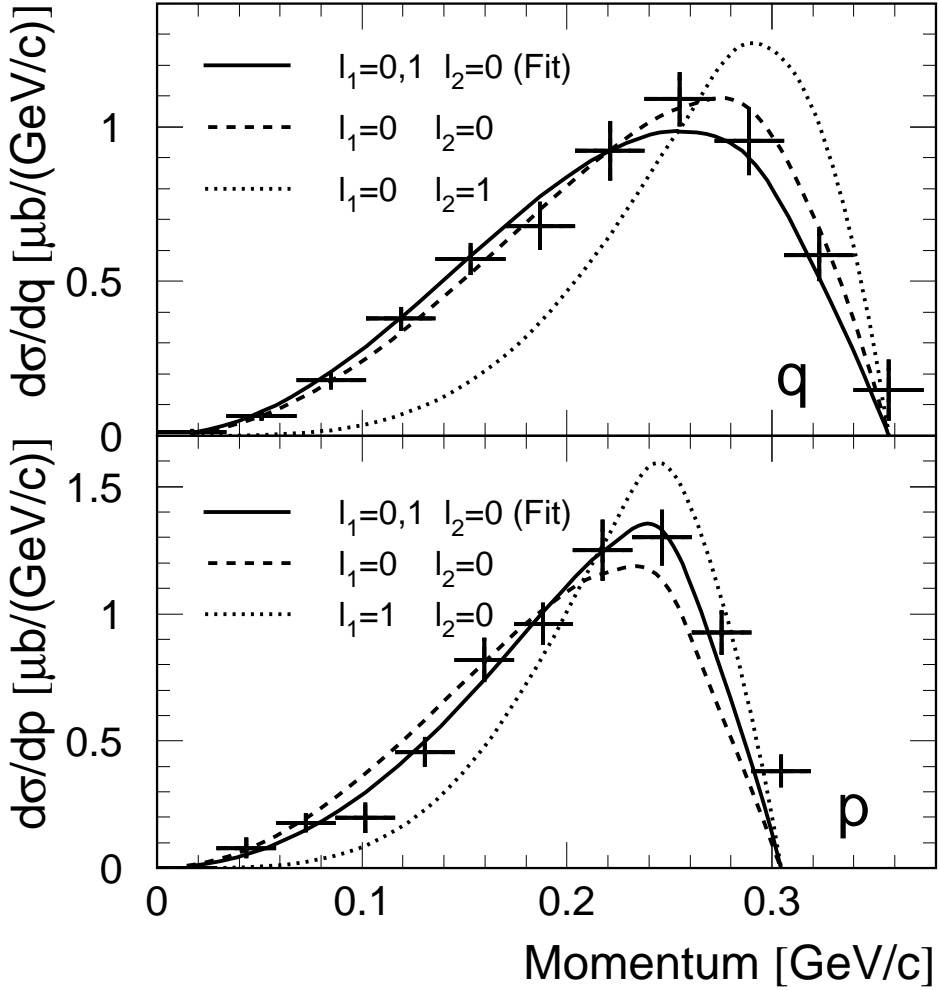,width=0.75\linewidth}
  \end{center}
  \caption{
    Differential cross section for the $pp\phi$ reaction as a function of the
    CM momentum $q$ of the $\phi$ meson (upper frame), and as a function of
    the proton momentum $p$ in the $pp$ reference frame (lower frame).  The
    dashed curves denote the behavior of the three body phase space when the
    $\phi$ meson is in a Ss wave state (i.e. $l_1=l_2=0$) relative to the
    protons.  The dotted curves correspond to the phase space distribution for
    $l_1=0, \ l_2=1$ (upper frame) and for $l_1=1, \ l_2=0$ (lower frame).
    The solid lines represent a simultaneous fit to both sets of data
    presented here with the combination of S and P wave contributions in the
    $pp$ exit channel.}
\label{fig:qpMom}
\end{figure}

  Assuming that the matrix elements have little variation across the available
  phase space, then the expected differential cross sections as a function of
  $q$ and $p$ are proportional to the variation of the three body phase space
  with $q$ and $p$.  To illustrate this, the $\phi$ meson differential cross
  section is plotted in the top frame of Figure~\ref{fig:qpMom} as a function
  of $q$.  The dashed curve shows the q dependence of the three body phase
  space for $l_1=l_2=0$, which was normalized to give the smallest $\chi^2$
  relative to the measured data. This curve describes the data well.  In
  contrast, the first moment of the dotted curve which denotes the case where
  the $\phi$ meson is in a P wave relative to the nucleons ($l_1=0, l_2=1$) is
  significantly higher than for the data.  These results are consistent with
  the observation from the data Figure~\ref{fig:1080} that the $\phi$ meson is
  in a nearly pure S wave state relative to the protons.
    
  Although the $l_1=l_2=0$ case describes the data well, a slightly better
  description of the data is generated by including P wave contributions in
  the proton-proton system (i.e.  $l_1 = 0, 1$ and $l_2 = 0$). The relative
  contributions of $l_1 = 0$ and $l_1 = 1$ has been determined by a combined
  fit (solid curve) to the data in the top frame of Figure~\ref{fig:qpMom}
  together with the differential cross section as a function of $p$, which is
  shown in the lower frame of Figure~\ref{fig:qpMom}.  The dashed curve in
  Figure~\ref{fig:qpMom} denotes the behavior of three body phase space for
  $l_1 = l_2 = 0$. In this case the data appear to have a significantly higher
  first moment of $p$, and as a result, the best description of the data
  includes a P wave contribution in the $pp$ system.  A simultaneous fit to
  the $d\sigma/dp$ and $d\sigma/dq$ data (solid curve) yields the following
  ratio of the mean matrix elements for the $l_1 = 1$, $l_2 = 0$ to the $l_1 =
  l_2 = 0$ states:
  \begin{equation}
\frac{|M_{10}|^2}{ |M_{00}|^2 + |M_{10}|^2} = 0.28 \pm 0.07.
  \label{Eq:M10/M00}
  \end{equation}
  
  Further confirmation that higher partial waves are involved in $\phi$ meson
  production at this beam momentum can be taken from the angular distribution
  of the daughter kaons from $\phi$ meson decay.  At threshold (i.e.
  $l_1=l_2=0$) the Pauli principle requires the outgoing protons to be in a
  $^1S_0$ state, and thus the total angular momentum and parity of the system
  must be $J^\pi=1^-$.  In this case, angular momentum and parity conservation
  require the $pp$ entrance channel to be in a $^3P_1$ state.  Since the
  orbital projection along the beam direction $m_L=0$ for the incident plane
  wave, the angular momentum coupling coefficients require the incident $pp$
  spin, and hence the outgoing $\phi$ meson spin, to be aligned along the
  beam axis~\cite{Ell95,Tit99,Rek97}. Consequently, the angular distribution
  of the daughter kaons in the $\phi$ meson reference frame must display a
  $\sin^2\Theta_\phi^K$ distribution relative to the beam direction.  At
  finite energies above threshold the spin alignment of the $\phi$ meson is
  diluted by contributions from higher partial waves, thereby modifying the
  expected angular distribution of the daughter kaons.
  
  The $\phi$ meson spin alignment can be quantified by the spin density
  matrix.  The elements of the spin density matrix are related to the emission
  angles of the kaons from the $\phi$ meson decay.  After integrating over the
  azimuthal emission angle ($\phi_\phi^K$) and imposing
  $\rho_{11}=\rho_{-1-1}$ and $\rho_{11} + \rho_{00} + \rho_{-1-1} = 1.0$, the
  diagonal elements of the spin density matrix are related by the following
  formula to the angular distribution of the daughter kaons (see
  e.g.~\cite{Tit99}):
  \begin{equation}
  W(\Theta_\phi^K) = \frac{3}{2} \left[ \rho_{11}\sin^2\Theta_\phi^K +
                                        \rho_{00}\cos^2\Theta_\phi^K \right].
  \label{Eq:dN/dQKf}
  \end{equation} 
  
\begin{figure}
  \begin{center}
    \epsfig{file=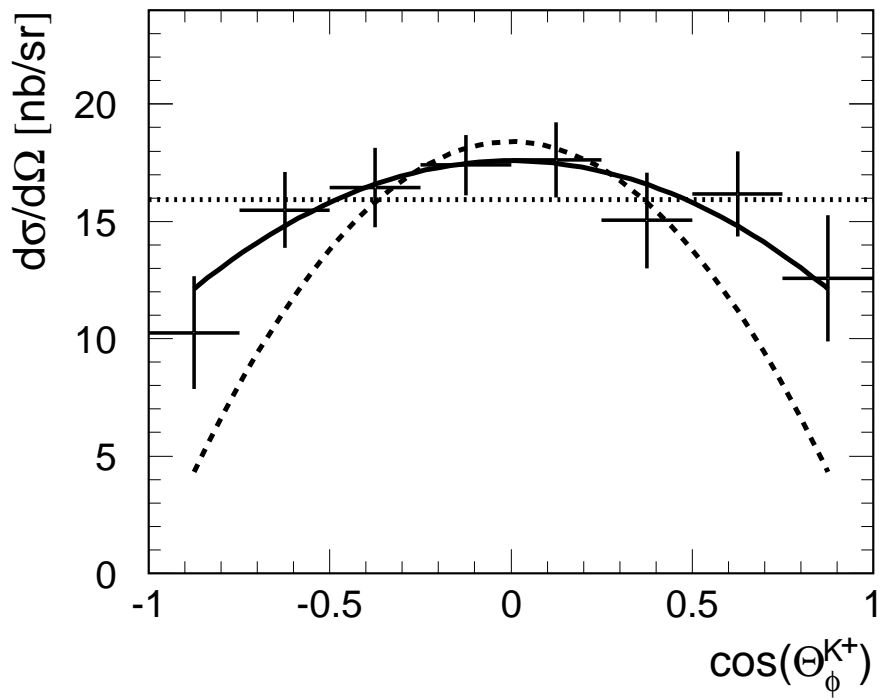,width=0.7\linewidth}
  \end{center}
  \caption{Differential cross section for the $pp\phi$ reaction plotted as 
    a function of the cosine of the polar angle of the daughter $K^+$ mesons
    measured in the $\phi$ meson reference frame, relative to the beam
    direction.  The curves represent different assumptions for the $\phi$
    meson spin alignment along the beam axis: zero alignment (dotted), full
    alignment as expected at threshold (dashed), and a fit to the data based
    on partial alignment (solid).}
  \label{fig:1090}
\end{figure}

  The differential cross section for the $pp\to pp\phi$ reaction has been
  evaluated and is presented in Figure~\ref{fig:1090} as a function of
  $\cos\Theta_\phi^K$, based on acceptance Matrix A described in
  Table~\ref{tab:KinVarPhiEtaOm}.  The dotted curve in this figure represents an
  isotropic distribution and the dashed curve corresponds to a
  $\sin^2\Theta_\phi^K$ distribution, which is the expected behavior at
  threshold due to the complete alignment of the $\phi$ meson spin (i.e.
  $\rho_{00}=0.0$).  Within the statistical errors, the measured data are not
  consistent with the $\sin^2\Theta_\phi^K$ distribution.  The solid curve is
  a fit to the data with Eq.~\ref{Eq:dN/dQKf}, from which the spin density
  matrix element $\rho_{00} = 0.23\pm0.04$ is determined.  The deviation of
  $\rho_{00}$ from the threshold value, together with the result presented in
  Eq.~\ref{Eq:M10/M00} and the lower frame of Figure~\ref{fig:70}, indicates a
  significant admixture of $^3P_{1,2}$ partial waves in the outgoing protons
  for $\phi$ meson production at this beam momentum.  ($^3P_0$ is forbidden in
  conjunction with $l_2=0$, since it would require a $1^+$ $pp$ entrance
  channel).
  
  Although the angular distribution shown in Figure~\ref{fig:1090} deviates
  from isotropy, the deviation is sufficiently small that it justifies as a
  reasonable approximation the neglect of $\Theta_\phi^K$ in acceptance Matrix
  B, used for the total cross section determination.

\section{Discussion}                           \label{sec:disc}

\begin{figure}
  \begin{center}
    \epsfig{file=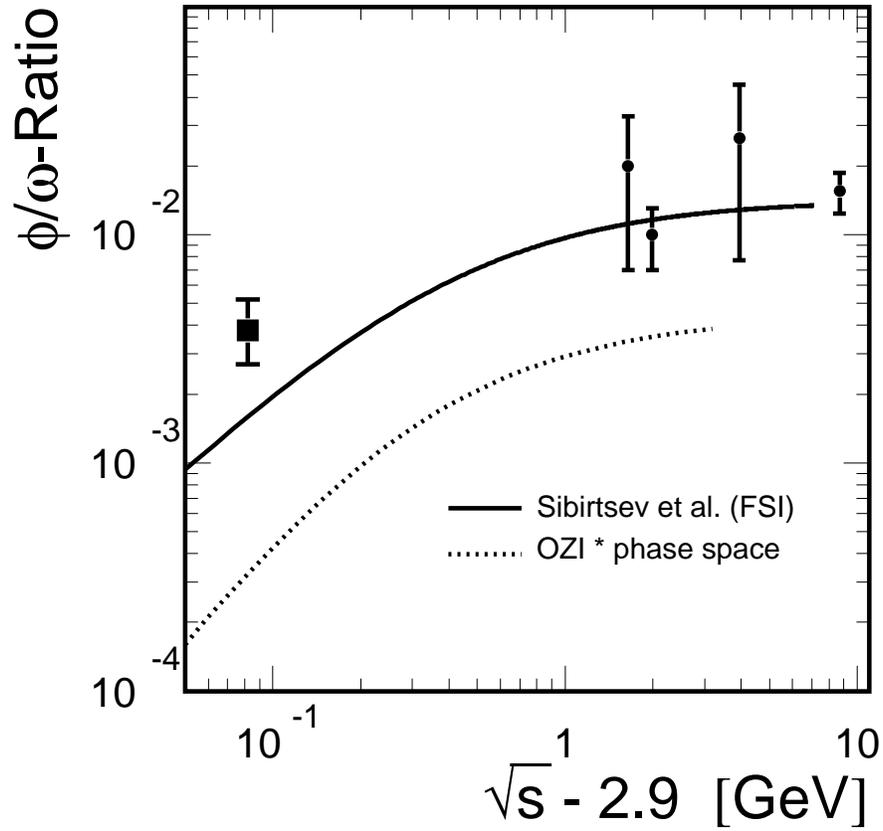,width=0.7\linewidth}
  \end{center}
  \caption {Ratio
    of the total cross sections for the $pp\phi$ and $pp\omega$ reactions as a
    function of the CM energy above the $\phi$ production threshold.
    Shown is the value measured in this work (square) together with data at
    higher energies and model calculations described in the text.}
  \label{fig:phi_om_rat}
\end{figure}

The ratio of the measured total cross sections for the $pp\to pp\phi$ and $pp\to
pp\omega$ reactions presented in this report is plotted as the filled square
in Figure~\ref{fig:phi_om_rat} in comparison to the other existing data at
higher energies~\cite{Blo75,Bal77,Are82,Gol97}.  This
data point is lower than the other data at higher energy.  This is primarily
due to the different mass of the $\phi$ and $\omega$ mesons, and the thereby
ensuing strong variation of the ratio of available three body phase space
volume near the $\phi$ meson threshold.  These data are compared to a
prediction based on a naive application of the OZI rule, including the
variation of the available phase space (dashed curve).  The data point from
this measurement is enhanced by roughly one order of magnitude relative to the
OZI prediction corrected for the available phase space volume.

The $\phi/\omega$ ratio presented here is based on cross sections measured at
the same beam momentum.  In order to reduce uncertainty related to the
different phase space volumes, partial wave amplitudes, and proton-proton
final state interactions, it is useful to compare the $\phi/\omega$ cross
section ratio at the same CM energy above threshold.  An evaluation based on
the solid curve in the lower frame of Figure~\ref{fig:eta_omega_xc} indicates
that the cross section for the $pp\to pp\omega$ reaction is about $8.5~\mu$b
at the same CM energy above threshold as for the $\phi$ meson in this
measurement (i.e.  Q=83~MeV).  In this case the $\phi/\omega$ ratio would only
be enhanced by about a factor 5 relative to the OZI rule, in agreement with
the higher energy data.

The solid curve in Figure~\ref{fig:phi_om_rat} is a calculation from Sibirtsev
et al.~\cite{Sib96} using a one pion exchange model and including the
proton-proton final state interaction.  These calculations, which describe the
higher energy data well, have an energy dependence similar to the ratio of
available phase space, and underestimate our point by about a factor of three.
Using $\pi-N$ data, Sibirtsev~\cite{Sib99b} extracts a ratio of the transition
amplitudes for $pp\omega$ to $pp\phi$ production $R = |M_\omega|/|M_\phi| =
8.5 \pm 1.0$.  Assuming that $R$ is independent of $\sqrt{s}$ near threshold,
this calculation predicts the $\phi$ meson cross section to be $(77 \pm
16)$~nb at $Q=83$~MeV.  A similar model by Chung et al.~\cite{Chu97} including
off-shell features of the pion and interference between the direct and
exchange diagrams predicts about 30~nb at this beam momentum.  After applying
the rather uncertain absolute normalization to our data (see
Section~\ref{ssec:absnorm}), these predictions can be compared to our measured
value of $190\pm14\pm80$~nb.

Another approach by Nakayama et al.~\cite{Nak99} explicitly includes not only
the mesonic current due to the $\pi\rho\to \phi$ coupling, but also the
nucleonic current where the $\phi$ meson couples directly to the nucleon.  The
observed angular distribution, which is nearly isotropic, indicates a
dominance of the mesonic current in contrast to the $\cos^2\Theta_{cm}^\phi$
distribution expected for the nucleonic current.  As a result of the dominance
of the mesonic current, they can not extract a unique value for the coupling
constant $g_{NN\phi}$.
In a similar model by Titov et al.~\cite{Tit99,Tit00}, our total cross section
and differential cross section as a function of $\Theta_{CM}^\phi$ can also be
reproduced without requiring an enhancement of $g_{\phi NN}$ over the OZI rule
prediction.  The importance of the correlated $\pi -\rho$ coupling, and the
small value of $g_{NN\phi}$ have both been predicted by Mei\ss ner et
al.~\cite{Mei97} to result from large cancellations between intermediate kaon
and hyperon graphs.
%
%
%

Although the value of $g_{\phi NN}$ is only poorly determined at this
point, due to uncertainties in the calculations it appears as though our
measured data can be explained without invoking a large explicit
violation of the OZI rule.
Nevertheless, the large decay width $\Gamma_{\phi\to \pi \rho}$
violates the OZI rule itself and requires kaon loop diagrams for a
quantitative explanation~\cite{Mar97}.  Thus, all these solutions require a
dominant role of intermediate states with open strangeness, thereby indicating
that there must indeed be a significant amount of strange sea quarks available
in reactions involving protons.

Ellis et al. claim that the enhanced $\phi$ meson production observed in $\bar
pp$ reactions proceeds dominantly through the ``rearrangement''
process~\cite{Ell99}.  To a large degree this is based on the observation that
the enhanced $\phi$ yield is strongly correlated to the initial spin triplet
state in $p\bar p$ annihilation~\cite{Ber96}.  Recent results on $\bar np$
annihilation in flight~\cite{Fil99} and $\bar pd$ Pontecorvo
reactions~\cite{Abe99} support this hypothesis.  According to their model,
the ``shake-out'' process should not depend upon the initial spin state, and
the ``rearrangement'' process should dominantly occur in the spin triplet
state for polarized strange sea quarks, in agreement with the data.
Following this argumentation, $\phi$ meson production in proton-proton
reactions is also expected to be strongly correlated with the spin triplet
initial state.

Directly at threshold, the proton-proton entrance channel must be in a $^3P_1$
state due to parity and angular momentum conservation, and consequently, the
spin of the $\phi$ meson is aligned along the beam axis. 
At $Q=83$~MeV we observe the spin density matrix
element $\rho_{00}$ to have a large deviation from the threshold
prediction.  This deviation is in qualitative agreement with the dilution of
the spin alignment expected due to the observed contribution of the $l_1=1$
partial wave in the exit channel, and suggests that a significant fraction
of the $\phi$ meson production at this beam momentum proceeds via the 
spin singlet $^1S_0$ and $^1D_2$ entrance channels.


\section{Summary and Conclusions}               \label{sec:sumcon}
In this paper total and differential cross section values for the production
of $\phi$ and $\omega$ mesons in proton-proton reactions at 3.67~GeV/c are
presented.  The total cross section ratio for these mesons is observed to be
about an order of magnitude larger than expected from predictions based on a
naive application of the OZI rule.  This enhancement is slightly larger than
the data measured at higher beam momenta, however significant uncertainty
remains regarding the relative contributions of different partial waves to the
$\phi$ versus $\omega$ production processes.

Apparent violations of the OZI rule in $\bar pp$ reactions have sometimes been
attributed to a significant contribution of intrinsic strangeness to the
proton's wave function.  On the other hand, most of the observed enhancement
of the $\phi$ meson yield in $\bar pp$ annihilation can be explained in terms
of rescattering and loop diagrams.  However, the large $\phi\to \pi\rho$
coupling also requires kaon loops, and thus the intermediate states are
dominated by hadrons with strange quark content.  Therefore, both
interpretations involve a significant contribution of strange sea quarks to
hadronic reactions involving protons.

Further information to help determine the origin of the strange sea quarks can
be taken from polarization observables.  For instance, based on a dominance of
the rearrangement process, the polarized intrinsic strange sea quarks prefer a
spin triplet initial state for $\phi$ meson production.  The differential
cross sections presented in this report indicate that the proton-proton
entrance channel is not in a pure $^3P_1$ state at the beam momentum of
3.67~GeV/c.  If the polarized intrinsic strangeness model is correct, then the
$\phi/\omega$ ratio should increase in direct proportion to the fraction of
spin triplet in the initial state.  Thus, it would be very useful to follow the
correlation of the $\phi/\omega$ ratio to the spin triplet fraction as a
function of beam momentum closer to threshold where the triplet fraction must
rise.

Another sensitive test of the intrinsic strangeness model would be to
determine the $\phi$ meson production cross section in proton-neutron
reactions.  For instance, based on the intrinsic strangeness model, Ellis et
al. predict the cross section ratio to be $\sigma_{np\to np\phi} /
\sigma_{pp\to pp\phi} \approx 0.25$ near threshold~\cite{Ell99}.  In contrast,
meson exchange models~\cite{Tit99,Rek97} predict
$\sigma_{np\to np\phi} / \sigma_{pp\to pp\phi} \approx 5$ near threshold.

Finally, it would also be very important to determine the $\omega$ meson yield
and partial wave contributions at the same excess energy relative to the
threshold as presented here for the $\phi$ meson.  The angular distributions
closer to threshold are needed to disentangle the mechanisms involved in the
$\omega$ meson production~\cite{Nak98}, and this would eliminate the
uncertainty related to the relative contributions of the different partial
waves in the $pp\omega$ and $pp\phi$ systems.

\section{Appendix: Data Tables}               \label{sec:data}

In this section the differential cross sections presented in the figures above
are listed in Tables~\ref{tab:omegaThetaCM},
\ref{tab:11070}, and \ref{tab:phiCMmom}.  In addition to the
statistical errors quoted here, there are the systematic errors and the overall
normalization uncertainty discussed above.

\acknowledgments

We would like to thank the staff of the Saturne Laboratory for providing
excellent experimental conditions.

This work has been supported in part by the following agencies:
CNRS-IN2P3, CEA-DSM, NSF, INFN, KBN (2 P03B 117 10 and 2 P03B 115 15)
and GSI.       

\noindent
$^a$ Present Address: DAPNIA/SPhN, CEA Saclay, France. \\
$^b$ Present Address: LPHNHE, Ecole Polytechnique 91128 Palaiseau, France.\\
$^c$ Present Address: Brokat Infosystems AG, Stuttgart, Germany. \\
$^d$ Present Address: Temple Univ., Philadelphia, Pennsylvania, USA \\
$^e$ Present Address: IU School of Medicine, Indianapolis, Indiana, USA.\\
$^f$ Deceased. \\
$^g$ Present Address: Motorola Polska Software Center, Krak\'ow, Poland.

\newpage

\begin{table}[h]
    \caption{Kinematic variables and the number of bins per
      variable associated with the acceptance correction matrices for the
      $ppK^+K^-$ final state (Matrices A and B), and for the $ppX$ final states,
      where $X=\eta, \omega$.} 
    \begin{tabular}{ccccccccc}
      Kinematic Variable &
      $\left( M_{inv}^{p_1X}\right)^2$ &
      $\left( M_{inv}^{p_2X}\right)^2$ &
      $\Theta_{CM}^X$ &
      $\phi_{CM}^X$ &
      $\Psi_{CM}^{pp}$ &
      $M_{inv}^{KK}$ &
      $\Theta_X^K$ &
      $\phi_X^K$  \\
\hline
      Matrix A ($X=K^+K^-$) & 4 & 4 & 1  & 1 & 4 & 1  & 10 & 4 \\
      Matrix B ($X=K^+K^-$) & 1 & 1 & 10 & 1 & 4 & 40 & 1  & 1 \\
      $pp\to ppX (X=\eta,\omega)$   & 10 & 10 & 20 & 1 &4 & -- & -- & -- \\
    \end{tabular}
    \label{tab:KinVarPhiEtaOm}
\end{table}


\begin{table}[h]
 \caption{List of systematic error sources and their effect for the
          various measured particle ratios.}
  \begin{tabular}{ccccc}
Systematic Error & $\phi/\omega$ & $\omega/\eta$ &$K^+K^-/\eta$& $\phi/\eta$\\
\hline
Acceptance Correction&  5\%       &  5\%          &  5\%        &  5\%       \\
$\omega$ Background  & 15\%       & 15\%          & --          &  --        \\
$\eta$ Background    & --         & 15\%          & 15\%        & 15\%       \\
$\phi$ Background    & 15\%       & --            & --          & 15\%       \\
Trigger Bias         & 10\%       & --            & 10\%        & 10\%       \\
Tracking Efficiency   & 10\%       & 10\%          & 10\%        & 10\%       \\
Drift in Electronics & 10\%       & 10\%          & 10\%        & 10\%       \\
$K^+ K^-$ Identification & 10\%   & --            & 10\%        & 10\%       \\
$\pi^+\pi^-$ Identification& 10\% & 5\%           & 10\%        & 10\%       \\
\hline
 Total               & 32\%       & 27\%          & 27\%        & 32\%       \\
  \end{tabular}
 \label{tab:uncert}
\end{table}

\newpage

\phantom{I do not understand where these figures will appear: Latex has a 
mind of its own...}

\begin{table}
 \caption{List of systematic biases and their effect for the
          various measured particle ratios.}
  \begin{tabular}{cccc}
Systematic Bias  & $\phi/\omega$ & $\omega/\eta$ & $\phi/\eta$   \\
\hline
Non-target Events      &  1.05         &  1.00         &  1.05         \\
Trigger Bias           &  0.93         &  1.00         &  0.93         \\
$\phi$ Identification  &  1.07         &  --           &  1.07         \\
$\omega$ Identification&  0.93         &  1.08         &  --           \\
$\eta$ Identification  &  --           &  0.94         &  0.94         \\ 
\hline
 Total                 &  0.97         &  1.02         &  0.98         \\
  \end{tabular}
 \label{tab:bias}
\end{table}

\begin{table}
\caption{Ratios of the total meson production cross
           sections for various reaction combinations including the
           corresponding statistical and systematic uncertainties. 
           N.B., the values and errors corresponding to the 
           $\phi/\eta$ and $\phi/\omega$ ratios have been 
           multiplied by $10^3$. }
\begin{tabular}{cccc}
                          & Ratio  & Statistical Error  & Systematic Error \\
\hline
$\omega/\eta$             & $0.37$ & $\pm0.02$          & $+0.1 -0.08$  \\
$\phi/\eta   \times 10^3$ & $1.42$ & $\pm0.1$           & $+0.45 -0.34$ \\
$\phi/\omega \times 10^3$ & $3.8 $ & $\pm0.2$           & $+1.2 -0.9$   \\
\end{tabular}
\label{tab:ratios}
\end{table}

\begin{table}
 \caption{Total production cross section for the reaction  
   $pp\rightarrow ppK^+K^-$ at 3.67~GeV/c and for the resonant ($\phi$ meson) 
   and non-resonant contributions, together with the statistical and
   systematic error, respectively.} 
  \begin{tabular}{cl}
   Production Channel           &  Cross Section [$\mu$b]      \\
\hline
   Non-resonant  $K^+K^-$  &  $0.11 \pm 0.009  \pm 0.046$  \\
   $\phi \rightarrow K^+K^- $
                           &  $0.09 \pm 0.007 \pm 0.04$    \\ 
   Total $K^+K^-$          &  $0.20 \pm 0.011  \pm 0.08 $  \\
\hline
   $(\phi \rightarrow K^+K^- ) \times \Gamma_{tot}/\Gamma_{K^+K^-}$
                           &  $0.19\pm 0.014 \pm 0.08$     \\
  \end{tabular}
 \label{tab:KKCross}
\end{table}

\newpage

\begin{table}
\caption{Differential cross sections for
$\omega$ (left two columns) and $\phi$ (right two columns) meson production
as functions of $\cos(\Theta_{CM}^\omega)$ and
$\cos(\Theta_{CM}^\phi)$, respectively.}
\begin{tabular}{ccccc}
$\cos(\Theta_{CM}^\omega)$&$d\sigma/d\Omega \ [\mu \mbox{b/sr}]$ & \hfill &
$\cos(\Theta_{CM}^\phi)$  &$d\sigma/d\Omega \ [\mbox{nb/sr}]$\\
\hline
-0.9  & ---              &&  -0.9  &  8.6  $\pm$ 29.5\\
-0.7  & 3.5   $\pm$ 2.0  &&  -0.7  & 15.3  $\pm$ 4.5 \\
-0.5  & 3.3   $\pm$ 0.4  &&  -0.5  & 17.4  $\pm$ 3.3 \\
-0.3  & 2.9   $\pm$ 0.34 &&  -0.3  & 16.6  $\pm$ 2.0 \\
-0.1  & 3.0   $\pm$ 0.29 &&  -0.1  & 17.2  $\pm$ 2.0 \\
 0.1  & 3.1   $\pm$ 0.23 &&   0.1  & 13.3  $\pm$ 1.5 \\
 0.3  & 2.9   $\pm$ 0.13 &&   0.3  & 16.6  $\pm$ 1.4 \\
 0.5  & 3.3   $\pm$ 0.07 &&   0.5  & 14.5  $\pm$ 1.2 \\
 0.7  & 3.9   $\pm$ 0.09 &&   0.7  & 15.4  $\pm$ 1.2 \\
 0.85 & 5.4   $\pm$ 0.1  &&   0.9  & 16.2  $\pm$ 1.2 \\
 0.95 & 8.0   $\pm$ 0.1  &&        & \\
\end{tabular}
\label{tab:omegaThetaCM}
\end{table}

\newpage

\begin{table}
 \caption{Differential cross sections for the $pp\to pp\phi$ reaction
  as a function of the angle of a proton measured in the $pp$ reference 
  frame relative to the beam direction (left two columns) and relative to 
  the direction of the $\phi$ meson (center two columns). The right two 
  columns tabulate the differential cross section as a function of the 
  polar angle of the daughter $K^+$ meson, 
  measured in the $\phi$ meson reference frame, relative to the beam
  direction. }
 \begin{tabular}{cccccccc}
 $\cos(\Theta_{pp}^p)$&$d\sigma/d\Omega \ [\mbox{nb/sr}]$&\hfill&
 $\cos(\Psi_{pp}^p)$  &$d\sigma/d\Omega \ [\mbox{nb/sr}]$&\hfill&
$\cos(\Theta_{\phi}^{K^+})$&$d\sigma/d\Omega \ [\mbox{nb/sr}]$\\
\hline
-0.875 & 15.4  $\pm$ 2.5 && -0.875 & 18.7  $\pm$ 2.7 && -0.875 & 10.3 $\pm$ 2.4\\
-0.625 & 16.2  $\pm$ 1.7 && -0.625 & 17.1  $\pm$ 3.8 && -0.625 & 15.5 $\pm$ 1.6\\
-0.375 & 15.3  $\pm$ 1.9 && -0.375 & 14.2  $\pm$ 2.8 && -0.375 & 16.4 $\pm$ 1.7\\
-0.125 & 14.8  $\pm$ 1.8 && -0.125 & 11.8  $\pm$ 2.4 && -0.125 & 17.4 $\pm$ 1.3\\
 0.125 & 15.9  $\pm$ 1.9 &&  0.125 & 12.2  $\pm$ 1.9 &&  0.125 & 17.6 $\pm$ 1.6\\
 0.375 & 14.1  $\pm$ 1.6 &&  0.375 & 14.4  $\pm$ 2.2 &&  0.375 & 15.0 $\pm$ 2.0\\
 0.625 & 15.8  $\pm$ 1.8 &&  0.625 & 14.5  $\pm$ 2.4 &&  0.625 & 16.2 $\pm$ 1.8\\
 0.875 & 13.5  $\pm$ 1.5 &&  0.875 & 18.0  $\pm$ 2.9 &&  0.875 & 12.6 $\pm$ 2.7\\
 \end{tabular}
\label{tab:11070}
\end{table}
         
\begin{table}
\caption{Differential  cross sections
  for the $pp\to pp\phi$ reaction as a function of the  CM momentum of the
  $\phi$ meson ($q$, left two columns) and as a function of the proton
  momentum in the proton-proton reference frame ($p$, right two columns).}
\begin{tabular}{ccccc}
q [GeV/c]  & $d\sigma/dq \ [\mu \mbox{b/(GeV/c)}]$ & \hfill &
p [GeV/c]  & $d\sigma/dp \ [\mu \mbox{b/(GeV/c)}]$\\
\hline
 0.017 & 0.011  $\pm$ 0.012 && 0.0145 & 0.0    $\pm$ 0.55  \\
 0.051 & 0.064  $\pm$ 0.015 && 0.0435 & 0.079  $\pm$ 0.040 \\
 0.085 & 0.180  $\pm$ 0.030 && 0.0725 & 0.178  $\pm$ 0.037 \\
 0.119 & 0.379  $\pm$ 0.038 && 0.1015 & 0.199  $\pm$ 0.060 \\
 0.153 & 0.573  $\pm$ 0.052 && 0.1305 & 0.457  $\pm$ 0.058 \\
 0.187 & 0.680  $\pm$ 0.078 && 0.1595 & 0.819  $\pm$ 0.087 \\
 0.221 & 0.923  $\pm$ 0.097 && 0.1885 & 0.961  $\pm$ 0.083 \\
 0.255 & 1.09   $\pm$ 0.087 && 0.2175 & 1.25   $\pm$ 0.12  \\
 0.289 & 0.955  $\pm$ 0.11  && 0.2465 & 1.30   $\pm$ 0.11  \\
 0.323 & 0.585  $\pm$ 0.083 && 0.2755 & 0.926  $\pm$ 0.087 \\
 0.357 & 0.148  $\pm$ 0.098 && 0.3045 & 0.382  $\pm$ 0.065 \\
\end{tabular}
\label{tab:phiCMmom}
\end{table}
\phantom{help}



\newpage

\phantom{it is a real pain trying to get the figures into Revtex}

\newpage

\newpage



\newpage



\newpage

\end{document}